\documentclass[aps, prd, superscriptaddress, preprintnumbers, floatfix, longbibliography,10pt,twocolumn]{revtex4-2}

\usepackage{graphicx,amsfonts,amsmath,amssymb,amstext}
\usepackage{float,wrapfig}
\usepackage{subfigure, psfrag}
\usepackage{dsfont,bm}
\usepackage{color}
\usepackage[colorlinks=true,urlcolor=blue,citecolor=blue,linkcolor=blue]{hyperref}
\hypersetup{breaklinks=true}
\usepackage{verbatim}
\usepackage[normalem]{ulem}

\newcommand{\RE}[1]{\,\mbox{Re}\left\{{#1}\right\}}

\usepackage{array}
\newcolumntype{L}[1]{>{\raggedright\let\newline\\\arraybackslash\hspace{0pt}}m{#1}}
\newcolumntype{C}[1]{>{\centering\let\newline\\\arraybackslash\hspace{0pt}}m{#1}}
\newcolumntype{R}[1]{>{\raggedleft\let\newline\\\arraybackslash\hspace{0pt}}m{#1}}

\definecolor{purple}{rgb}{0.8,0,0.6}
\definecolor{darkgreen}{rgb}{0.00,0.6,0.00}

\usepackage[usenames,dvipsnames]{xcolor}

\extrafloats{100}

\begin{document}

\title{Optical conductivity of bilayer dice lattices}
\date{August 31, 2023}

\author{P.~O.~Sukhachov}
\email{pavlo.sukhachov@yale.edu}
\affiliation{Department of Physics, Yale University, New Haven, Connecticut 06520, USA}

\author{D.~O.~Oriekhov}
\affiliation{Instituut-Lorentz, Universiteit Leiden, P.O. Box 9506, 2300 RA Leiden, The Netherlands}

\author{E.~V.~Gorbar}
\affiliation{Faculty of Physics, Kyiv National Taras Shevchenko University, 64/13 Volodymyrska st., 01601 Kyiv, Ukraine}
\affiliation{Bogolyubov Institute for Theoretical Physics, 14-b Metrolohichna st., 03143 Kyiv, Ukraine}

\begin{abstract}

We calculate optical conductivity for bilayer dice lattices in commensurate vertically aligned stackings. The interband optical conductivity reveals a rich activation behavior unique for each of the four stackings. We found that the intermediate energy band, which corresponds to the flat band of a single-layer dice lattice, plays a different role for different stackings. The interband selection rules, which are crucial for the single-layer lattice, may become lifted in bilayer lattices. The results for effective and tight-binding models are found to be in qualitative agreement for some of the stackings and the reasons for the discrepancies for others are identified. Our findings propose optical conductivity as an effective tool to distinguish between different stackings in bilayer dice lattices.
\end{abstract}

\maketitle

\section{Introduction}
\label{sec:Introduction}

Optical response provides a powerful way to extract a wealth of information about different properties of materials. The sensitivity to interband transitions distinguishes optical or alternating current response from its direct current counterpart. By tuning the frequency of electromagnetic radiation, one can probe different interband transitions, determine the selection rules, and map the energy bands of various materials including those with unusual spectra such as graphene as well as Weyl and Dirac semimetals.

Perhaps, the most distinct feature of the interband optical response in 2D Dirac materials with relativistic-like energy spectrum is the presence of the steplike feature originating from the Pauli principle followed by a plateau; see Refs.~\cite{Gusynin:2005iv,Gusynin-Carbotte:2006,Nair-Geim-graphene:2008,Li-Basov-graphene:2008} for theoretical and experimental studies of the optical response in graphene. In 3D Dirac and Weyl semimetals, the plateau is replaced with linearly growing bulk~\footnote{Due to the interplay of the bulk and surface contributions, the optical response in 3D materials is more involved compared to the 2D ones. For example, even under the conditions of the normal skin effect, the penetration and reflection of electromagnetic radiation from Weyl and Dirac semimetals subject to external magnetic fields can be unusual~\cite{Sukhachov-Glazman:2021-skin,Matus-Surowka:2021}.} interband optical conductivity as long as the radiation frequency is sufficiently high to surpass the Pauli blocking; see, e.g., Refs.~\cite{Burkov-Balents:2011,Hosur-Vishwanath:2012,Rosenstein:2013,Ashby-Carbotte:2014-opt,Neubauer-Pronin-Cd3As2:2016,Jenkins-Drew:2016,Wu-Orenstein-TaAs:2016,Xu-Qiu:2016-TaAs-optics}. A more detailed review of the results for the optical conductivity in nodal metals can be found in Refs.~\cite{Armitage:rev-2018,Pronin-Dressel:rev-2020,GMSS:book}.

Recently, materials with even more exotic and complex spectra containing flat bands started to attract significant attention. In 2D, the flat band energy spectrum can be realized by twisting layers of bilayer graphene~\cite{LopesdosSantos-CastroNeto:2007,Morell-Barticevic:2010,Bistritzer-MacDonald:2010}; alternatively, flat bands can occur in certain lattices such as the dice ($\mathcal{T}_3$) lattice~\cite{Sutherland:1986,Vidal-Doucot:1998}. The dice lattice has a hexagonal structure with an additional atom placed in the center of each hexagon. The central atom represents a hub that connects to six rims. The rims form two sublattices where each of the rims connects to three hubs. Since there are three atoms per unit cell, the energy spectrum of a dice lattice contains three bands and can be represented as a Dirac point intersected by a flat band~\cite{Raoux-Montambaux:2013}. Such a crossing point is described in terms of spin-1 fermions; several bands crossing at the same point might allow for higher-spin fermions. While, to the best of our knowledge, there are no solid-state materials realizing dice lattices, the latter were proposed in artificial systems such as optical lattices~\cite{Rizzi-Fazio:2005,Bercioux-Haeusler:2009} and Josephson arrays~\cite{Serret-Pannetier:2002}. Other types of 2D lattices that produce flat bands include kagome~\cite{Syozi:1951} and Lieb~\cite{Lieb:1989} lattices; see Ref.~\cite{Leykam-Flach:2018} for a review of artificial flat-band systems.

The unusual energy spectrum of higher-spin fermions and crossing points is directly manifested in optical responses where additional interband transitions can become possible and the overall scaling of the optical conductivity with frequency can change. The interband transitions involving flat bands are manifested as a steplike feature with the activation frequency equal to the Fermi energy~\cite{Illes-Nicol:2016} followed by a plateau in the optical conductivity. This behavior is similar to that in graphene, where, due to the absence of the flat band, the activation frequency is different and is equal to the double Fermi energy. For 2D spin-1 and higher-spin fermions, optical and magneto-optical conductivities were calculated in Refs.~\cite{Malcolm-Nicol:2014,Illes-Nicol:2016,Biswas-Ghosh:2016,Kovacs-Cserti:2016,Illes-Nicol:2016b,Iurov-Huang:2018,Chen-Ma:2019,Carbotte-Nicol:2019,Iurov-Huang:2020,Han-Lai:2022,Oriekhov-Gusynin:2022,Iurov-Huang:2022,Tamang-Biswas:2022} and plasmon excitations were studied in Refs.~\cite{Malcolm-Nicol:2016,Balassis-Roslyak:2019,Iurov-Huang:2020,Iurov-Weekes:2021,Iurov-Abranyos:2022,Han-Lai:2022}. In 3D, similar multi-fold energy spectra with spin-1 and even higher-spin fermions were proposed in Ref.~\cite{Bradlyn-Bernevig:2016}. Experimentally, multifold fermions were realized in chiral materials such as CoSi~\cite{Takane-Sato:2019,Rao-Ding:2019,Sanchez-Hasan:2018}, RhSi~\cite{Sanchez-Hasan:2018}, and AlPt~\cite{Schroter-Chen:2019}. The optical conductivity of 3D higher-spin fermions was studied in Refs.~\cite{Flicker-Grushin:2018,Sanchez-Martinez-Grushin:2019,Habibi-Jafari:2019,Xu-Grushin-Wu:2020}. A natural intermediate step between 2D and 3D systems is to consider a few-layer system, which, as bilayer graphene, might reveal a different set of properties, see Ref.~\cite{Abergel-Falko:2006} for the optical conductivity of bilayer graphene. We introduced such bilayer dice lattices and studied their energy spectra in Ref.~\cite{SOG:part1-2023}.

In this work, we investigate the optical conductivity in tight-binding and effective models derived in our work~\cite{SOG:part1-2023} for the four nonequivalent commensurate stackings of bilayer dice lattices with vertically aligned atoms: (i) aligned $AA-BB-CC$, (ii) hub-aligned $AB-BA-CC$, (iii) mixed $AA-BC-CB$, and (iv) cyclic $AB-BC-CA$; here, $A$ and $B$ denote rim sites and $C$ denotes hub sites. We found that the activation behavior and scaling of the optical conductivity drastically depend on the stacking. The obtained results for the effective and tight-binding models are in qualitative agreement for certain stackings if the interband transitions for the low-frequency optical response are saturated by the three-band-crossing points. Discrepancies between the effective and tight-binding approaches appear if the energy spectrum away from the band-crossing points also contributes to the interband transitions. In this case, we identify the dominant optical transitions and show that the optical conductivity is dominated by the local extrema of the dispersion relation. We found also that the intermediate bands play no role in the effective models for the hub-aligned $AB-BA-CC$ and mixed $AA-BC-CB$ stacking, which allowed us to use the particle-hole-asymmetric semi-Dirac and tilted Dirac models. Such a reduction is not possible for the aligned $AA-BB-CC$ and cyclic $AB-BC-CA$ stackings where all three bands contribute to the optical conductivity. While the aligned $AA-BB-CC$ stacking inherits the optical selection rules of the single-layer lattice, i.e., only the transitions involving the flat band are allowed, all bands may contribute to the optical conductivity for other stackings. The obtained in this work results provide an effective way to distinguish different stackings of dice lattices in optical responses.

The paper is organized as follows. We summarize the tight-binding and effective low-energy models of bilayer dice lattices in Sec.~\ref{sec:Model}. The optical conductivity for each of the four nonequivalent stackings is discussed in Sec.~\ref{sec:Conductivity}. The results are summarized in Sec.~\ref{sec:Summary}. Technical details concerning the non-abbreviated effective models and the calculation of the optical conductivity are given in Appendices~\ref{sec:App-Models-full} and \ref{sec:App-Conductivity}, respectively. Throughout this paper, we use $k_B=1$.

\section{Model}
\label{sec:Model}

In this section, by following Ref.~\cite{SOG:part1-2023}, we summarize the tight-binding and effective Hamiltonians of bilayer dice lattices in commensurate stackings.

\subsection{Tight-binding models}
\label{sec:Model-TB}

\begin{figure*}[t]
\centering
\includegraphics[width=0.24\textwidth]{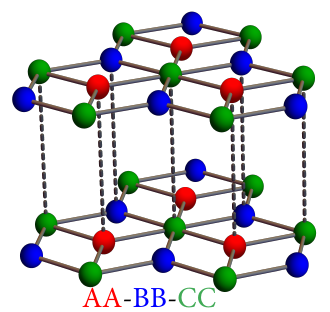}
\includegraphics[width=0.24\textwidth]{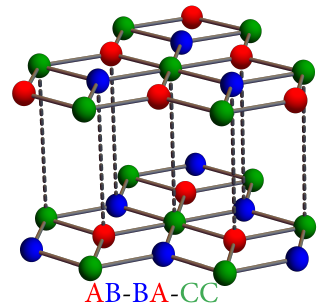}
\includegraphics[width=0.24\textwidth]{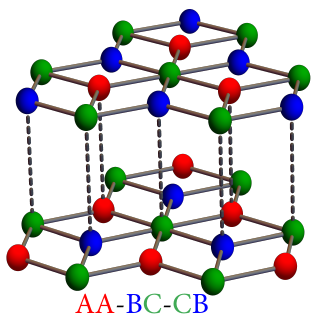}
\includegraphics[width=0.24\textwidth]{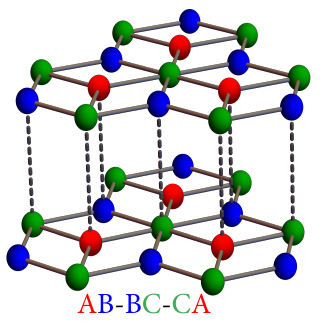}
\caption{The schematic representation of bilayer dice lattice in (i) aligned $AA-BB-CC$, (ii) hub-aligned $AB-BA-CC$, (iii) mixed $AA-BC-CB$, and (iv) cyclic $AB-BC-CA$ stackings. The $A$, $B$, and $C$ sites are denoted by red, blue, and green points.}
\label{fig:Model-dice}
\end{figure*}

In the basis of states corresponding to the $A$, $C$, and $B$ sublattices, the tight-binding Hamiltonian of single-layer dice lattices is~\cite{Raoux-Montambaux:2013}
\begin{equation}
\label{Model-tb-h-def}
H_{0}(\mathbf{q}) = \left(
                  \begin{array}{ccc}
                    0 & -t \sum_{j}e^{-i\mathbf{q}\cdot\bm{\delta}_j} & 0 \\
                    -t \sum_{j}e^{i\mathbf{q}\cdot\bm{\delta}_j} & 0 & -t \sum_{j}e^{-i\mathbf{q}\cdot\bm{\delta}_j} \\
                    0 & -t \sum_{j}e^{i\mathbf{q}\cdot\bm{\delta}_j} & 0 \\
                  \end{array}
                \right),
\end{equation}
where $t$ is the hopping constant, $\mathbf{q}$ is the wave vector in the Brillouin zone, and the vectors $\bm{\delta}_1 = a \left\{0,1\right\}$, $\bm{\delta}_2 = a \left\{\sqrt{3},-1\right\}/2$, and $\bm{\delta}_3 = -a \left\{\sqrt{3},1\right\}/2$ denote the relative positions of the sites $A$ (rims) with respect to the sites $C$ (hubs). The parameter $a$ determines the distance between neighboring $A$ and $C$ sites. Sites $B$ (rims) are related to sites $A$ by the $C_3$ rotational symmetry with respect to sites $C$; the structure of a single-layer dice lattice can be also seen in each of the layers of bilayer lattices shown in Fig.~\ref{fig:Model-dice}. The energy spectrum of the Hamiltonian~(\ref{Model-tb-h-def}) resembles that in graphene and reveals two nonequivalent Dirac points in the hexagonal Brillouin zone. However, the Dirac points are intersected by a zero-energy flat band.

Let us now discuss bilayer dice lattices. The corresponding tight-binding Hamiltonian is defined as
\begin{equation}
\label{Model-H-total}
H(\mathbf{q}) = \left(
                  \begin{array}{cc}
                    H_{0}(\mathbf{q}) & H_{\rm c} \\
                    H_{\rm c}^{\rm T} & H_{0}(\mathbf{q}) \\
                  \end{array}
                \right),
\end{equation}
where $H_{0}(\mathbf{q})$ is the single-layer tight-binding Hamiltonian (\ref{Model-tb-h-def}) and $H_{\rm c}$ describes the inter-layer coupling. As we proposed in Ref.~\cite{SOG:part1-2023}, there are four nonequivalent commensurate stackings for a bilayer dice lattice with vertically aligned sites: (i) aligned $AA-BB-CC$, (ii) hub-aligned $AB-BA-CC$, (iii) mixed $AA-BC-CB$, and (iv) cyclic $AB-BC-CA$. The bilayer dice lattices for these stackings are shown in Fig.~\ref{fig:Model-dice}. Assuming only nearest-neighbor tunneling and, for simplicity, equal tunneling strength for all sites, we use the following coupling Hamiltonians $H_{\rm c}$ connected with the aligned, hub-aligned, mixed, and cyclic stackings, respectively:
\begin{eqnarray}
\label{Model-h-tun-1-2-3-4}
&&H_{\rm c}^{\rm (a)} = g \left(
                  \begin{array}{ccc}
                    1 & 0 & 0 \\
                    0 & 1 & 0 \\
                    0 & 0 & 1 \\
                  \end{array}
                \right), \quad
                H_{\rm c}^{\rm (h)} = g \left(
                  \begin{array}{ccc}
                    0 & 0 & 1 \\
                    0 & 1 & 0 \\
                    1 & 0 & 0 \\
                  \end{array}
                \right), \nonumber\\
                &&H_{\rm c}^{\rm (m)} = g \left(
                  \begin{array}{ccc}
                    1 & 0 & 0 \\
                    0 & 0 & 1 \\
                    0 & 1 & 0 \\
                  \end{array}
                \right), \quad  H_{\rm c}^{\rm (c)} = g \left(
                  \begin{array}{ccc}
                    0 & 0 & 1 \\
                    1 & 0 & 0 \\
                    0 & 1 & 0 \\
                  \end{array}
                \right).
\end{eqnarray}
Here, $g\geq0$ is the coupling strength.

The tight-binding Hamiltonian (\ref{Model-H-total}) supplemented with the corresponding coupling Hamiltonian (\ref{Model-h-tun-1-2-3-4}) defines the spectral and transport properties of a dice bilayer lattice. However, its relatively high dimension ($6\times 6$) and intricate structure complicate the analysis. Therefore, to make an analytical advance and to develop physical intuition, we employ effective low-energy models valid in the vicinity of the threefold band-crossing $K$ (or $K^{\prime}$) points. In what follows, we summarize the corresponding effective Hamiltonians. The details of the derivation and the energy spectrum can be found in Ref.~\cite{SOG:part1-2023}; see also Figs.~\ref{fig:Conductivity-aligned-TB-full}(b), \ref{fig:Conductivity-hub-TB-full}(b), \ref{fig:Conductivity-mixed-TB-full}(b), and \ref{fig:Conductivity-cyclic-TB-full}(b).

\subsection{Effective models}
\label{sec:Model-effective}

We start our discussion of the effective models with the simplest, \emph{aligned $AA-BB-CC$}, stacking. The effective Hamiltonian in the vicinity of the $K$ point is
\begin{equation}
\label{Model-effective-aligned-Heff}
H_{\rm eff}^{\rm (a)} = g \mathds{1}_3 + \hbar v_F \left(\mathbf{S}\cdot \mathbf{k}\right),
\end{equation}
where the momentum $\mathbf{k} = \mathbf{q} - 4\pi \left\{1,0\right\}/(3\sqrt{3}a)$ is measured with respect to the $K$ point,
\begin{equation}
\label{Model-S-def}
S_x = \frac{1}{\sqrt{2}}\left(
        \begin{array}{ccc}
          0 & 1 & 0 \\
          1 & 0 & 1 \\
          0 & 1 & 0 \\
        \end{array}
      \right) \quad  \mbox{and} \quad
      S_y = \frac{1}{\sqrt{2}} \left(
        \begin{array}{ccc}
          0 & -i & 0 \\
          i & 0 & -i \\
          0 & i & 0 \\
        \end{array}
      \right)
\end{equation}
are the (pseudo)spin-1 matrices, and $v_F = 3ta/(\sqrt{2}\hbar)$ is the Fermi velocity. In the leading nontrivial order in $\hbar v_F k/g$, the effective model for the aligned stacking is represented by two copies of the single-layer linearized Hamiltonians separated by $2g$ in energy; the Hamiltonian for the other copy is obtained by replacing $g\to-g$ in Eq.~(\ref{Model-effective-aligned-Heff}). The energy spectrum contains flat and two dispersive branches: $\epsilon_{0} = g$, $\epsilon_{1} = g + \hbar v_F k$, and $\epsilon_{2} = g - \hbar v_F k$.

The abbreviated effective Hamiltonian for the \emph{hub-aligned $AB-BA-CC$} stacking reads
\begin{eqnarray}
\label{Model-effective-hub-Heff}
H_{\rm eff}^{\rm (h)} &=& g \mathds{1}_3 + \frac{\hbar v_F}{\sqrt{2}} k_x \left(
                                                            \begin{array}{ccc}
                                                              0 & 1 & 0 \\
                                                              1 & 0 & 1 \\
                                                              0 & 1 & 0 \\
                                                            \end{array}
                                                          \right) \nonumber\\
                                                          &+&\left(\frac{\hbar v_F}{\sqrt{2}} \right)^2 \frac{k_y^2}{2g}\left(
                                                                     \begin{array}{ccc}
                                                                       1 & 0 & -1 \\
                                                                       0 & 2 & 0 \\
                                                                       -1 & 0 & 1 \\
                                                                     \end{array}
                                                                   \right).
\end{eqnarray}
Compared to the effective Hamiltonian in Ref.~\cite{SOG:part1-2023}, we omitted a few terms quadratic in the wave vector which are not crucial for the qualitative shape of the spectrum and, as we will demonstrate in Sec.~\ref{sec:Conductivity-aligned}, do not affect the main features of the optical conductivity; for the sake of completeness, the nonabbreviated effective model is given in Eq.~(\ref{App-Model-effective-hub-Heff}). The energy spectrum of Hamiltonian (\ref{Model-effective-hub-Heff}) in the vicinity of the $K$ point is
\begin{eqnarray}
\label{Model-effective-hub-eps-0}
\epsilon_0  &=& g + \frac{(\hbar v_Fk_y)^2}{2g},\\
\label{Model-effective-hub-eps-1}
\epsilon_{1} &=& g + \frac{(\hbar v_Fk_y)^2}{4g} + \hbar v_F \sqrt{k_x^2 +\left(\frac{\hbar v_F}{4g}\right)^2 k_y^4},\\
\label{Model-effective-hub-eps-2}
\epsilon_{2} &=&  g + \frac{(\hbar v_Fk_y)^2}{4g} - \hbar v_F \sqrt{k_x^2 +\left(\frac{\hbar v_F}{4g}\right)^2k_y^4}.
\end{eqnarray}
The above energy spectrum corresponds to a particle-hole-asymmetric version of the semi-Dirac model~\cite{Hasegawa-Kohmoto:2006} in which the dispersion relation is linear in one direction and quadratic in the other. The particle-hole asymmetry around the band-crossing points is quantified by the momentum-dependent $\sim (\hbar v_Fk_y)^2/g$ term.

In the case of the \emph{mixed $AA-BC-CB$} stacking, the abbreviated effective Hamiltonian reads
\begin{equation}
\label{Model-effective-mixed-Heff}
H_{\rm eff}^{\rm (m)} = g \mathds{1}_3 + \frac{\hbar v_F}{2\sqrt{2}} \left(
                                                            \begin{array}{ccc}
                                                              0 & 2k_x & k_{-} \\
                                                              2k_x & 0 & k_{-} \\
                                                              k_{+} & k_{+} & 0 \\
                                                            \end{array}
                                                          \right),
\end{equation}
where $k_{\pm}=k_x\pm ik_y$. Quadratic terms are important for the additional energy branch where they describe its anisotropy and introduce a dependence on $k_y$. However, as we will show in Sec.~\ref{sec:Conductivity-hub}, this additional branch does not play any role in the interband transitions for the effective model. The energy spectrum of Hamiltonian (\ref{Model-effective-mixed-Heff}) reads
\begin{eqnarray}
\label{Model-effective-mixed-eps-0}
\epsilon_0 &=& g - \frac{\hbar v_F}{\sqrt{2}} k_x, \\
\label{Model-effective-mixed-eps-1}
\epsilon_1 &=& g + \frac{\hbar v_F}{2\sqrt{2}} k_x + \frac{\hbar v_F}{2\sqrt{2}} \sqrt{3k_x^2 +2k_y^2}, \\
\label{Model-effective-mixed-eps-2}
\epsilon_2 &=& g + \frac{\hbar v_F}{2\sqrt{2}} k_x - \frac{\hbar v_F}{2\sqrt{2}} \sqrt{3k_x^2 +2k_y^2}.
\end{eqnarray}

Finally, the effective linearized Hamiltonian for the \emph{cyclic $AB-BC-CA$} stacking is
\begin{equation}
\label{Model-effective-cyclic-Heff}
H_{\rm eff}^{\rm (c)} = g \mathds{1}_3 + \frac{\hbar v_F}{2\sqrt{2}} \left(
                                                            \begin{array}{ccc}
                                                              0 & k_{-} & k_{+} \\
                                                              k_{+} & 0 & 2k_{-} \\
                                                              k_{-} & 2k_{+} & 0 \\
                                                            \end{array}
                                                          \right).
\end{equation}
Its energy spectrum is
\begin{equation}
\label{Model-effective-cyclic-eps-0-1-2}
\epsilon_n = g + \hbar v_F k \cos{\left\{\frac{1}{3} \mbox{arccos}{\left[\frac{\cos{(3\varphi)}}{\sqrt{2}}\right]} -\frac{2\pi (1-n)}{3}\right\}},
\end{equation}
where $n=0,1,2$ and, to simplify the expressions, we used the polar coordinate system with $\left\{k_x, k_y\right\} = k \left\{\cos{(\varphi)}, \sin{(\varphi)}\right\}$.

\section{Optical conductivity}
\label{sec:Conductivity}

In this section, we calculate optical conductivity for the commensurate stackings of the bilayer dice lattice described in Sec.~\ref{sec:Model}. Optical conductivities for each of the four stackings are presented in Secs.~\ref{sec:Conductivity-aligned}--\ref{sec:Conductivity-cyclic}, respectively. The results for effective models are analyzed and compared with those in the tight-binding models.

\subsection{Kubo linear response approach}
\label{sec:Conductivity-Kubo}

Let us start with formulating the linear response approach. The optical conductivity tensor is defined in terms of the retarded current-current correlation function
\begin{equation}
\label{Conductivity-Kubo-sigma-def}
\sigma_{nm}(\Omega) =  -i\frac{\hbar}{\Omega} \Pi_{nm}(\Omega+i0; \mathbf{0}),
\end{equation}
where $\Omega$ is the frequency of the oscillating electromagnetic field and the polarization tensor is given by
\begin{eqnarray}
\label{Conductivity-Kubo-Pi-def}
\Pi_{nm}(\Omega +i0;\mathbf{0}) &=&  e^2 \frac{T}{\hbar} \sum_{l=-\infty}^{\infty} \int \frac{d^2 k}{(2\pi)^2} \mbox{tr} \Big[ v_n G(i\omega_l; \mathbf{k}) \nonumber\\
&\times& v_m G(i\omega_l-\Omega -i0; \mathbf{k})\Big] \nonumber\\
&=& -e^2 \int\int  d\omega  d \omega^{\prime} \frac{f^{\rm eq}(\hbar\omega)-f^{\rm eq}(\hbar\omega^{\prime})}{\omega -\omega^{\prime}-\Omega-i0} \nonumber\\
&\times& \int \frac{d^2 k}{(2\pi)^2} \mbox{tr} \left[ v_n A(\omega; \mathbf{k}) v_m A(\omega^\prime; \mathbf{k})\right]. \nonumber\\
\end{eqnarray}
Here, $\omega_l =(2l+1)\pi T/\hbar$ is the fermion Matsubara frequency, $l$ is an integer, $T$ is temperature in energy units, and $v_n =  \partial_{k_n} H(\mathbf{k})/\hbar$ is the velocity matrix.
The Green function in the momentum space reads
\begin{equation}
\label{Conductivity-G-def}
G(\omega \pm i 0; \mathbf{k})  = \frac{i}{\hbar \omega -\mu -H(\mathbf{k}) \pm i0},
\end{equation}
where $\mu$ is the chemical potential and the signs $\pm$ correspond to the retarded ($+$) and advanced ($-$) Green functions. In the last expression in Eq.~(\ref{Conductivity-Kubo-Pi-def}), we performed the summation over Matsubara frequencies as well as introduced the Fermi-Dirac distribution function $f^{\rm eq}(\epsilon)= 1/\left[e^{(\epsilon-\mu)/T}+1\right]$ and the spectral function
\begin{equation}
\label{Conductivity-A-def}
A(\omega; \mathbf{k}) =\frac{1}{2\pi} \left[ G(\omega+i 0; \mathbf{k}) -G(\omega-i 0; \mathbf{k}) \right] \Big|_{\mu=0}.
\end{equation}

The calculation of the real part of the conductivity tensor can be significantly simplified if the trace in Eq.~(\ref{Conductivity-Kubo-Pi-def}) is real. Then, by using the identity
\begin{equation}
\label{Conductivity-Kubo-P-value}
\frac{1}{\omega -\omega^\prime-\Omega \mp i 0} = \mbox{p.v.}\frac{1}{\omega -\omega^\prime-\Omega}
\pm i \pi \delta\left(\omega -\omega^\prime-\Omega\right),
\end{equation}
one can straightforwardly extract the imaginary part of $\Pi_{nm}(\Omega+i0,\mathbf{0})$. Here, $\mbox{p.v.}$ stands for the principal value.

For the diagonal part of the conductivity, the trace in Eq.~(\ref{Conductivity-Kubo-Pi-def}) is real; see also Appendix~\ref{sec:App-Conductivity} for explicit calculations. Therefore, we have the following expression for $\mbox{Re}{\left\{\sigma_{nn}(\Omega)\right\}}$:
\begin{eqnarray}
\label{Conductivity-Kubo-sigma-nn}
\mbox{Re}{\left\{\sigma_{nn}(\Omega)\right\}} &=& -\frac{\pi \hbar e^2}{\Omega} \int  d\omega \left[f^{\rm eq}(\hbar\omega)-f^{\rm eq}(\hbar\omega -\hbar \Omega)\right] \nonumber\\
&\times&\int \frac{d^2 k}{(2\pi)^2} \mbox{tr} \left[ v_n A(\omega; \mathbf{k}) v_n A(\omega-\Omega; \mathbf{k})\right]. \nonumber\\
\end{eqnarray}
The imaginary part can be derived via the Kramers-Kronig relations; see, e.g., Ref.~\cite{Han-Lai:2022} for the corresponding calculations in a single-layer dice lattice. As for the off-diagonal components, $\mbox{Re}{\left\{\sigma_{n m}(\Omega)\right\}}$ with $n\neq m$, their absence is guaranteed by the time-reversal symmetry.

The expression for the conductivity in Eq.~(\ref{Conductivity-Kubo-sigma-nn}) is valid both for effective and tight-binding models, as well as contains intra- and interband terms. The intra-band part is nonuniversal and strongly depends on quasiparticle scattering mechanisms. Therefore, in our calculations for the effective models, we focus only on the interband part. In addition, we dispense with the effects of nonvanishing temperature and consider only the case $T\to0$.

To identify the contributions of different bands in the optical conductivity, it is convenient to use the following Kubo-Greenwood formula~\cite{Mahan:book-2013} at vanishing temperature:
\begin{eqnarray}
\label{Conductivity-Kubo-sigma-nn-TB}
\mbox{Re}{\left\{\sigma_{xx}(\Omega)\right\}}\!\! &=&\! \pi\! \sum_{s\neq s^{\prime}} \sum_{\mathbf{k}} \frac{\theta\left(\mu-\epsilon_{s^{\prime}}\right)-\theta\left(\mu-\epsilon_{s}\right)}{\epsilon_{s}-\epsilon_{ s^{\prime}}} \nonumber\\
&\times&\!\! \delta\left(\hbar \Omega+\epsilon_{s}-\epsilon_{ s^{\prime}}\right)\left|\left\langle\Psi_s(\mathbf{k})\left|j_x(\mathbf{k}) \right| \Psi_{s^{\prime}}(\mathbf{k})\right\rangle\right|^2. \nonumber\\
\end{eqnarray}
Here $s,s^{\prime}=\pm \left\{2,0,1\right\}$ label energy bands with the overall sign corresponding to the triplets of the bands crossing at $\pm g$, respectively. The current operator is defined as $j_x(\mathbf{k}) = -e\partial_{k_x}H(\mathbf{k})/\hbar$ and $\Psi_s(\mathbf{k})$ are the eigenstates of $H(\mathbf{k})$. The conductivity tensor is isotropic in the tight-binding model, $\sigma_{xx}(\Omega)=\sigma_{yy}(\Omega)$. In our numerical calculations, we replace the $\delta$ function in Eq.~(\ref{Conductivity-Kubo-sigma-nn-TB}) by a Lorentzian of the half-width $\Gamma$; this is equivalent to replacing $i0\to i\Gamma$ in Eq.~(\ref{Conductivity-G-def}). The summation over momenta is performed over the Brillouin zone using a uniform discretization.

\subsection{Aligned \texorpdfstring{$AA-BB-CC$}{AA-BB-CC} stacking}
\label{sec:Conductivity-aligned}

Exploiting the fact that the effective Hamiltonian~(\ref{Model-effective-aligned-Heff}) for the aligned $AA-BB-CC$ stacking is equivalent (except the shifted position of the band-crossing point quantified by the coupling strength $g$) to its counterpart for the single-layer dice lattice, the final result for the optical conductivity $\mbox{Re}{\left\{\sigma_{xx}(\Omega)\right\}}$ summed over the $K$ and $K^{\prime}$ crossing points reads
\begin{equation}
\label{Conductivity-aligned-sigma-xx-fin}
\mbox{Re}{\left\{\sigma_{xx}(\Omega)\right\}} =
\sigma_0 \left[\theta\left(\hbar \Omega +g-\mu\right) -\theta\left(g-\mu -\hbar \Omega\right)\right],
\end{equation}
where $\sigma_0 = e^2/(4\hbar)$ and $\theta(x)$ is the unit step function; see Appendix~\ref{sec:App-Conductivity-aligned} for details. The first term in the last expression in Eq.~(\ref{Conductivity-aligned-sigma-xx-fin}) corresponds to the transitions from the flat band $\epsilon_{0}=g$ to the upper linear band $\epsilon_{1}=g + v_Fk$. The second term describes the transitions between the lower band $\epsilon_{2}=g - v_Fk$ and the flat band $\epsilon_{0}=g$. Notice that there are no direct transitions between the upper and lower dispersive bands. The obtained results for the effective model agree with those for the single-layer dice lattice where the direct transitions between the dispersive bands are also forbidden~\cite{Illes-Nicol:2016,Iurov-Huang:2020,Han-Lai:2022}.

For comparison, we present also the optical conductivity of monolayer graphene~\cite{Gusynin:2005iv,Gusynin-Carbotte:2006},
\begin{equation}
\label{Conductivity-aligned-sigma-xx-Graphene}
\mbox{Re}{\left\{\sigma_{xx}^{\rm (graphene)}(\Omega)\right\}} = \frac{\sigma_0}{2} \theta\left(\frac{\hbar \Omega}{2} -\mu \right).
\end{equation}
As one can see by comparing Eqs.~(\ref{Conductivity-aligned-sigma-xx-fin}) and (\ref{Conductivity-aligned-sigma-xx-Graphene}), the additional zero-energy band for the dice lattice allows for a different activation behavior where the steplike feature occurs at $\hbar \Omega=|\mu-g|$ rather than $\hbar \Omega=2\mu$.

\begin{figure*}[t]
\centering
\subfigure[]{\includegraphics[width=0.45\textwidth]{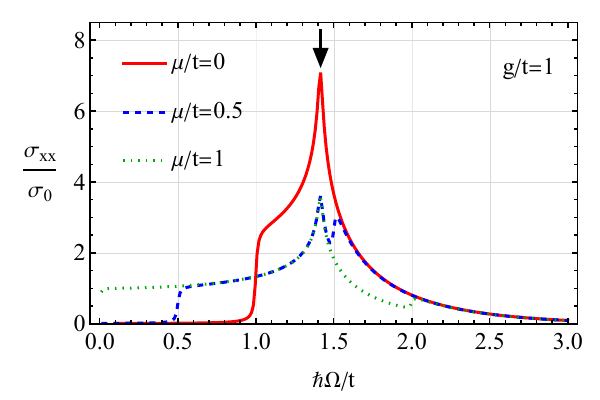}}
\hspace{0.01\textwidth}
\subfigure[]{\includegraphics[width=0.45\textwidth]{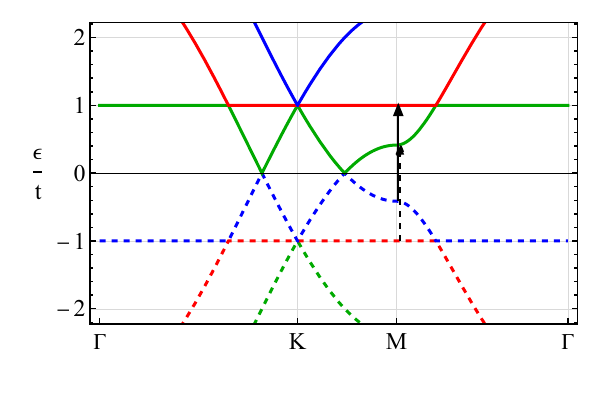}}
\caption{
Panel (a): The interband optical conductivity for the tight-binding Hamiltonian corresponding to the aligned $AA-BB-CC$ stacking at $\mu/t=0$ (solid red line), $\mu/t=0.5$ (dashed blue line), and $\mu/t=1$ (dotted green line). The black arrow marks the peak of the optical conductivity; the corresponding transition at $\mu/t=0$ is shown in panel (b) by the vertical black arrow. We used the phenomenological broadening $\Gamma/t = 0.01$.
Panel (b): The energy spectrum of the tight-binding Hamiltonian along the $\Gamma-\mbox{K}-\mbox{M}-\Gamma$ line in the Brillouin zone. The vertical black arrows mark the transitions corresponding to the peak in panel (a). In both panels, $\sigma_0=e^2/(4\hbar)$ and $g/t=1$.
}
\label{fig:Conductivity-aligned-TB-full}
\end{figure*}

To illustrate the role of other parts of the band structure away from the band-crossing points, we show the optical conductivity in the tight-binding model for a broader range of frequencies and Fermi energies in Fig.~\ref{fig:Conductivity-aligned-TB-full}. The appearance of the steplike feature at $\hbar \Omega = |\mu-g|$ agrees well with the result for the effective model; see Eq.~(\ref{Conductivity-aligned-sigma-xx-fin}). The steplike feature for $\mu=0$ is two times higher at the onset than that for, e.g., $\mu/t=0.5$, which is explained by the contributions of both flat bands with the energies $\pm g$, i.e., $\epsilon_{\pm 0}$. As in the effective model, the optical conductivity in the tight-binding one is saturated by the transitions between the dispersive and flat bands. In agreement with Eq.~(\ref{Conductivity-aligned-sigma-xx-fin}), the steplike feature at $\hbar\Omega = g$ is split into two steps at $|\mu+g|$ and $|\mu-g|$ if $\mu\neq0$. There is also a peak at large frequencies $\hbar\Omega/g \gtrsim 1$, see the vertical arrow in Fig.~\ref{fig:Conductivity-aligned-TB-full}(a). This peak corresponds to the transitions between the local extrema of the low-energy dispersive (flat) band $\epsilon_{-1}$ ($\epsilon_{-0}$) and the high-energy flat (dispersive) band $\epsilon_{+0}$ ($\epsilon_{+2}$) near the $M$ point; see dashed blue (red) and solid red (green) lines, respectively, in Fig.~\ref{fig:Conductivity-aligned-TB-full}(b). The description of such a feature is, of course, beyond the range of applicability of the effective model.

\subsection{Hub-aligned \texorpdfstring{$AB-BA-CC$}{AB-BA-CC} stacking}
\label{sec:Conductivity-hub}

The conductivity for the hub-aligned $AB-BA-CC$ stacking can be straightforwardly calculated by using both the tight-binding model and the effective Hamiltonian (\ref{Model-effective-hub-Heff}); see Appendices~\ref{sec:App-Conductivity-hub} and \ref{sec:App-Conductivity-SD} for details.

\begin{figure*}[t]
\centering
\subfigure[]{\includegraphics[width=0.45\textwidth]{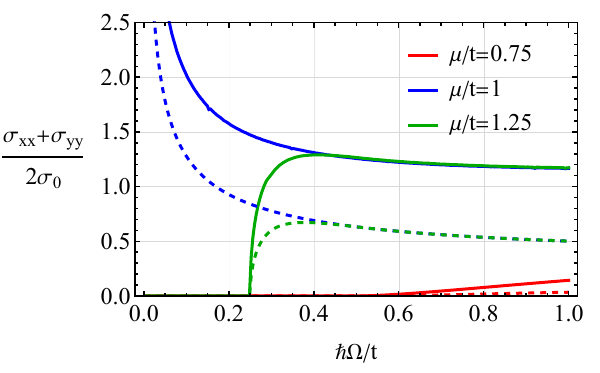}}
\hspace{0.01\textwidth}
\subfigure[]{\includegraphics[width=0.45\textwidth]{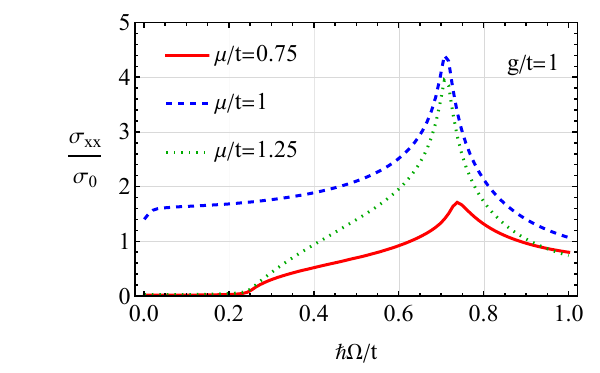}}
\caption{
The dependence of the averaged over all band-crossing points interband conductivity $\left(\RE{\sigma_{xx}}+\RE{\sigma_{yy}}\right)/2$ for the effective (panel (a)) and tight-binding (panel (b)) models of the hub-aligned $AB-BA-CC$ stacking at a few values of $\mu/t$. Solid and dashed lines in panel (a) correspond to the nonabbreviated and abbreviated effective models, respectively; see Eqs.~(\ref{App-Model-effective-hub-Heff}) and (\ref{Conductivity-hub-SD-H-def}), respectively. We used the phenomenological broadening $\Gamma/t=0.01$ in panel (b).
In both panels, $g/t=1$ and $\sigma_0 = e^2/(4\hbar)$.
}
\label{fig:Conductivity-hub-sigma-xx-yy-effective-TB}
\end{figure*}

To start with, we focus on the contribution of the band-crossing point (i.e., the $K$ or $K^{\prime}$ point) in the optical conductivity. We compare the results for the effective and tight-binding models in Fig.~\ref{fig:Conductivity-hub-sigma-xx-yy-effective-TB}. Since the conductivity for the tight-binding model takes into account all crossing points and is isotropic, we compare averaged conductivities $(\sigma_{xx}+\sigma_{yy})/2$. (In the tight-binding model  $\sigma_{xx}=\sigma_{yy}$.) As one can see, there is a noticeable difference between the conductivities for the effective and tight-binding models. Among the common features, we identify only the onsets of the conductivities for some Fermi energies. The rest of the profile is dominated by features of the energy spectrum away from the crossing points that are not captured by the effective model; see also the discussion below and Fig.~\ref{fig:Conductivity-hub-TB-full}. From the analysis of the effective model in Appendix~\ref{sec:App-Conductivity-SD}, we conclude that the onset frequency $\Omega_{\rm on}$, which is evident from Fig.~\ref{fig:Conductivity-hub-sigma-xx-yy-effective-TB}(a), is determined by the minimal distance between empty states at the $\epsilon_{+1}$ branch and filled states at the $\epsilon_{+2}$ branch. Therefore, unlike the aligned stacking, transitions between dispersing bands are allowed.

\begin{figure*}[t]
\centering
\subfigure[]{\includegraphics[width=0.45\textwidth]{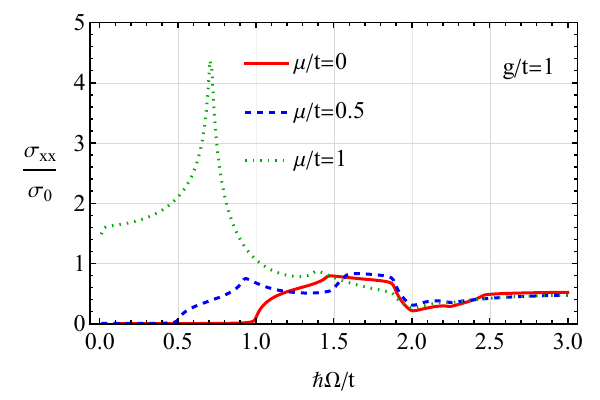}}
\hspace{0.01\textwidth}
\subfigure[]{\includegraphics[width=0.45\textwidth]{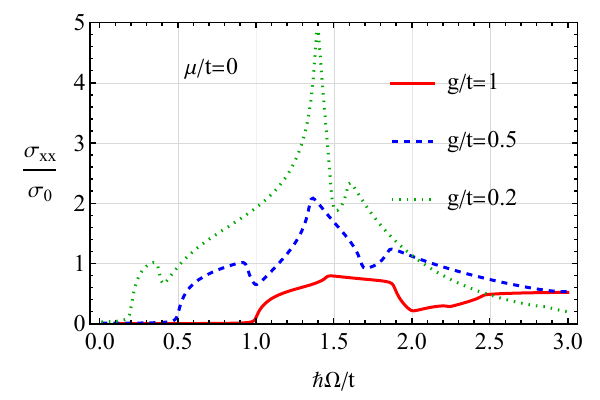}}
\caption{
The interband optical conductivity for the tight-binding model of the hub-aligned $AB-BA-CC$ stacking for a few values of $\mu$ at $g/t=1$ (panel (a)) and a few values of $g$ at $\mu=0$ (panel (b)). In both panels, $\sigma_0=e^2/(4\hbar)$ and we used the phenomenological broadening $\Gamma/t = 0.01$.
}
\label{fig:Conductivity-hub-TB-mu-g}
\end{figure*}

For Fermi energies away from the crossing points or at small coupling constants $g/t\lesssim 1$, the effective model is not applicable and we resort to the tight-binding one. We present the optical conductivity in the tight-binding model for a wider range of chemical potentials and coupling strengths in Figs.~\ref{fig:Conductivity-hub-TB-full} and \ref{fig:Conductivity-hub-TB-mu-g}. The nontrivial band structure for the hub-aligned $AB-BA-CC$ stacking leads to a few interesting features.  There are noticeable peaks at $\hbar\Omega/t \approx1.5$ for $\mu=0$ which are determined by the transitions between the dispersive $\epsilon_{-1}$ ($\epsilon_{+2}$) bands and intermediate $\epsilon_{+0}$ ($\epsilon_{-0}$); see Fig.~\ref{fig:Conductivity-hub-TB-full}(a). As one can see from Fig.~\ref{fig:Conductivity-hub-TB-full}(b), the peak appears due to the local extrema of the dispersion relation near the $M$ point in the Brillouin zone; the onset of the optical conductivity is determined by the transitions between the flatlike and dispersive bands, e.g., $\epsilon_{-0}$ and $\epsilon_{+2}$. Transitions between other bands, e.g., $\epsilon_{-2}\to \epsilon_{+2}$ and $\epsilon_{-1}\to \epsilon_{+1}$, are non-negligible only for high frequencies $\hbar \Omega \lesssim g$ and lead to a much smaller peak.

\begin{figure*}[t]
\centering
\subfigure[]{\includegraphics[width=0.45\textwidth]{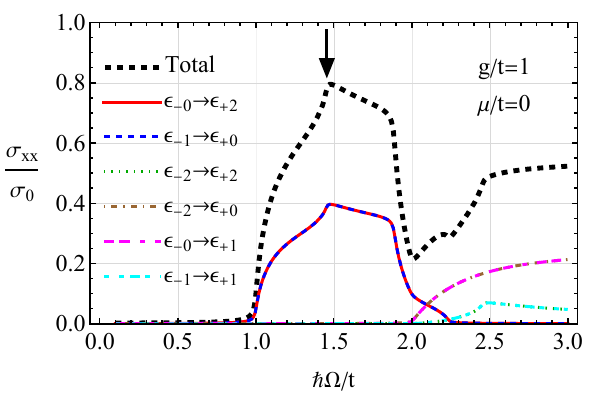}}
\hspace{0.01\textwidth}
\subfigure[]{\includegraphics[width=0.45\textwidth]{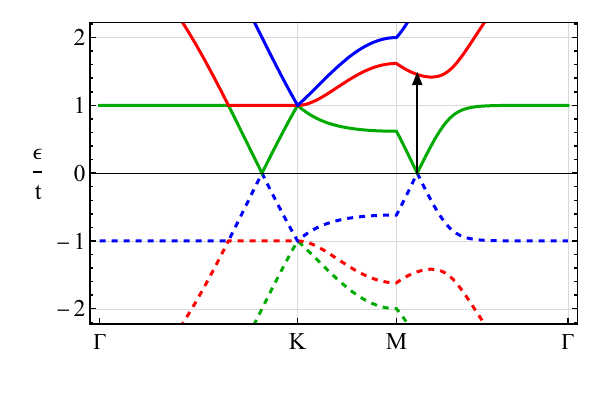}}
\caption{
Panel (a): The interband optical conductivity for the tight-binding model corresponding to the hub-aligned $AB-BA-CC$ stacking at $\mu=0$. The contributions due to different bands are marked by different lines. The vertical arrow marks the peak of the optical conductivity; the corresponding transitions are shown in panel (b). We use the phenomenological broadening $\Gamma/t = 0.01$.
Panel (b): The energy spectrum of the tight-binding Hamiltonian along the $\Gamma-\mbox{K}-\mbox{M}-\Gamma$ line in the Brillouin zone. The vertical arrow shows transitions contributing to the peaks in the optical conductivity at $\mu/t=0$. In both panels, $\sigma_0=e^2/(4\hbar)$ and $g/t=1$.
}
\label{fig:Conductivity-hub-TB-full}
\end{figure*}

As follows from Fig.~\ref{fig:Conductivity-hub-TB-mu-g}(a), the peak at $\hbar\Omega/t \approx 1.5$ is split and shifts to smaller frequencies with the rise of $\mu$. The peak at a smaller frequency $\hbar\Omega/t \lesssim 1$ corresponds to the transitions between $\epsilon_{+2}$ and $\epsilon_{+0}$. Its counterpart for $\epsilon_{-1} \to \epsilon_{+0}$ remains approximately at the same frequency. The peak at $\hbar\Omega/t \approx 1.5$ is split into three peaks and becomes more pronounced for smaller coupling constants $g$; see Fig.~\ref{fig:Conductivity-hub-TB-mu-g}(b). In agreement with our previous discussion and the results for the effective models, the onset frequency decreases since the bands shift to smaller frequencies at smaller $g$.

\subsection{Mixed \texorpdfstring{$AA-BC-CB$}{AA-BC-CB} stacking}
\label{sec:Conductivity-mixed}

\begin{figure*}[t]
\centering
\subfigure[]{\includegraphics[width=0.45\textwidth]{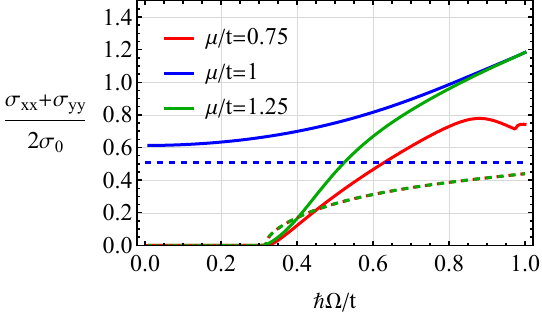}}
\hspace{0.01\textwidth}
\subfigure[]{\includegraphics[width=0.45\textwidth]{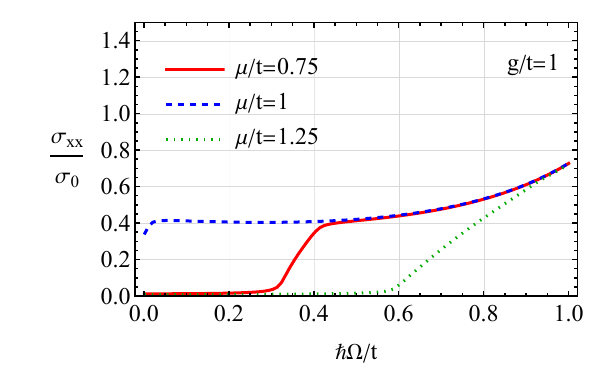}}
\caption{
The dependence of the averaged over all band-crossing points interband conductivity $\left(\RE{\sigma_{xx}}+\RE{\sigma_{yy}}\right)/2$ for the effective (panel (a)) and tight-binding (panel (b)) models of the mixed $AA-BC-CB$ stacking at a few values of $\mu/t$. Solid and dashed lines correspond to the non-abbreviated and abbreviated effective models, respectively; see Eqs.~(\ref{App-Model-effective-hub-Heff}) and (\ref{Conductivity-hub-SD-H-def}) for the definitions of the models. We used the phenomenological broadening $\Gamma/t=0.01$ in panel (b). In both panels, $\sigma_0 = e^2/(4\hbar)$ and $g/t=1$.
}
\label{fig:Conductivity-mixed-sigma-xx-yy-effective-TB}
\end{figure*}

Let us address the optical conductivity for the mixed $AA-BC-CB$ stacking. We start with investigating the role of the band-crossing points in optical conductivity. We compare the averaged optical conductivity obtained in the nonabbreviated and abbreviated effective models with the conductivity calculated in the tight-binding model in Fig.~\ref{fig:Conductivity-mixed-sigma-xx-yy-effective-TB}. The steplike dependence of the conductivity with the subsequent growth in Fig.~\ref{fig:Conductivity-mixed-sigma-xx-yy-effective-TB}(b) agrees with that for the effective model albeit only for certain Fermi energies: the onset frequencies for $\mu>g$ and $\mu<g$ are different in the tight-binding model. This is related to the contributions of other parts of the spectrum away from the crossing points.

\begin{figure*}[t]
\centering
\subfigure[]{\includegraphics[width=0.45\textwidth]{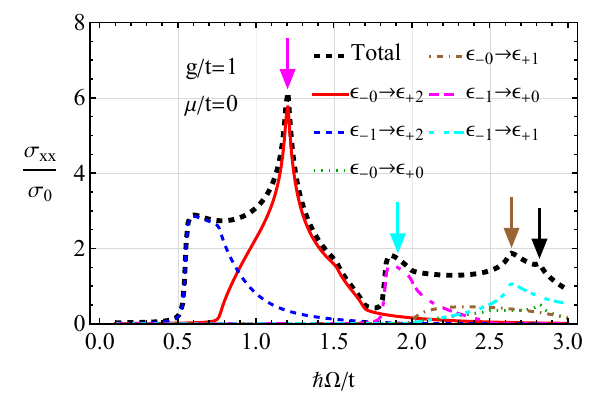}}
\hspace{0.01\textwidth}
\subfigure[]{\includegraphics[width=0.45\textwidth]{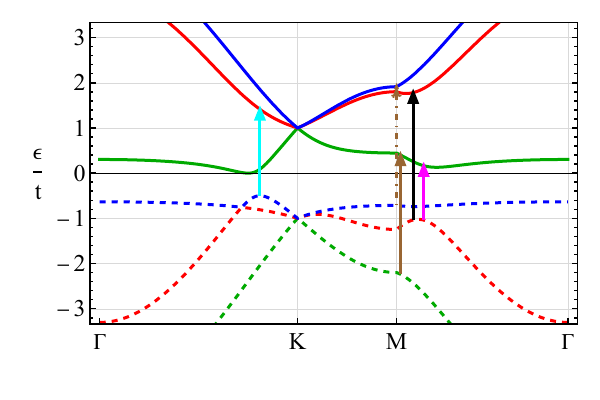}}
\caption{
Panel (a): The interband optical conductivity for the tight-binding model of the mixed $AA-BC-CB$ stacking at $\mu=0$. The contributions due to different bands are marked by different lines. Vertical arrows mark the peaks of the optical conductivity; the corresponding transitions are shown in panel (b). We used the phenomenological broadening $\Gamma/t = 0.01$.
Panel (b): The energy spectrum of the tight-binding Hamiltonian along the $\Gamma-\mbox{K}-\mbox{M}-\Gamma$ line in the Brillouin zone. The vertical arrows show the transitions contributing to the peaks in the optical conductivity at $\mu=0$, see panel (a).
In both panels, $\sigma_0=e^2/(4\hbar)$ and $g/t=1$.
}
\label{fig:Conductivity-mixed-TB-full}
\end{figure*}

The contributions of each of the bands to the optical conductivity are shown in Fig.~\ref{fig:Conductivity-mixed-TB-full}(a) and the corresponding transitions are marked in Fig.~\ref{fig:Conductivity-mixed-TB-full}(b). As one can see, while the onset is determined by the transitions between the upper occupied and the lowest empty bands, i.e., $\epsilon_{-1}\to \epsilon_{+2}$ at $\mu=0$, the most-pronounced peak originates from the transitions between the local extrema of the $\epsilon_{-0}$ and $\epsilon_{+2}$ bands near the $M$ point. Extrema for other bands near the $M$ point also lead to peaks albeit at higher frequencies and with smaller magnitudes.

\begin{figure*}[t]
\centering
\subfigure[]{\includegraphics[width=0.45\textwidth]{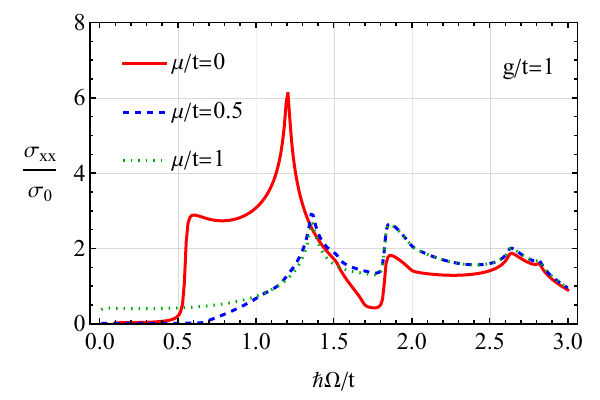}}
\hspace{0.01\textwidth}
\subfigure[]{\includegraphics[width=0.45\textwidth]{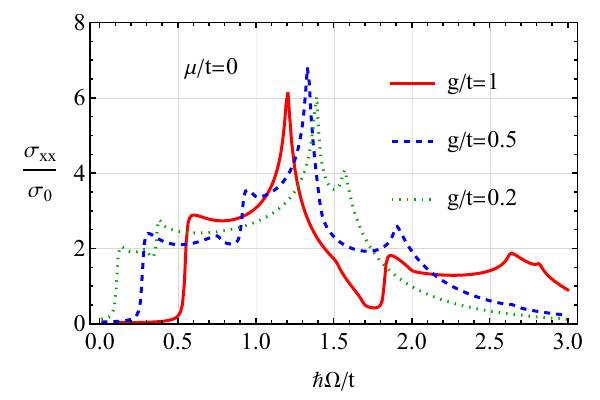}}
\caption{
The interband optical conductivity for the tight-binding model of the mixed $AA-BC-CB$ stacking for a few values of $\mu$ at $g/t=1$ (panel (a)) and a few values of $g$ at $\mu=0$ (panel (b)). In both panels, $\sigma_0=e^2/(4\hbar)$ and we used the phenomenological broadening $\Gamma/t = 0.01$.
}
\label{fig:Conductivity-mixed-TB-contributions}
\end{figure*}

The optical conductivity in the tight-binding model for several Fermi energies and coupling constants is shown in Fig.~\ref{fig:Conductivity-mixed-TB-contributions}. With the rise of the Fermi energy, low-frequency features become suppressed since the corresponding transitions are Pauli-blocked. The decrease of the coupling constant $g$ leads to the shift of the onset of the transitions to smaller frequencies but moves the central peaks to slightly higher frequencies. Indeed, the former is determined by the minimal distance between energy levels, which decreases at smaller $g$, while the latter originates from the extrema of the band structure near the $M$ point, which move away from each other at smaller $g$. The overall profile of the conductivity remains similar for different values of $g$.

\subsection{Cyclic \texorpdfstring{$AB-BC-CA$}{AB-BC-CA} stacking}
\label{sec:Conductivity-cyclic}

\begin{figure*}[t]
\centering
\subfigure[]{\includegraphics[width=0.45\textwidth]{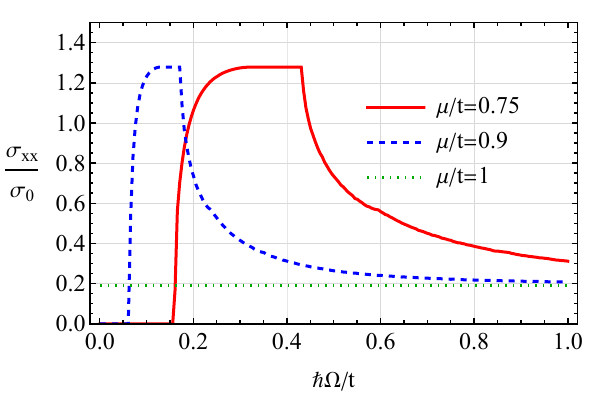}}
\hspace{0.01\textwidth}
\subfigure[]{\includegraphics[width=0.45\textwidth]{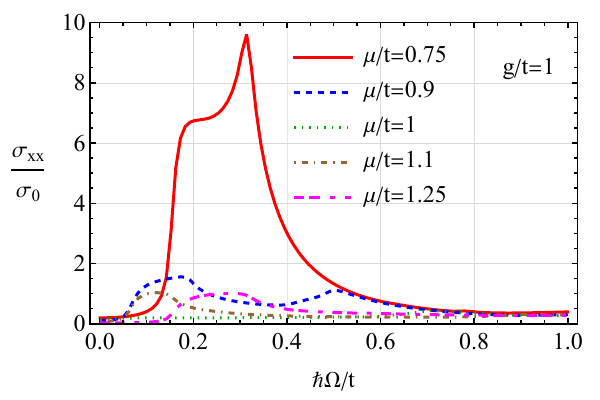}}
\caption{
Panel (a): The dependence of the interband conductivity $\RE{\sigma_{xx}}$ normalized to $\sigma_0 = e^2/(4\hbar)$ on $\hbar \Omega/t$ for the cyclic $AB-BC-CA$ stacking at a few values of $\mu/t$.
Panel (b): The interband conductivity for the tight-binding model of the same stacking where we used the phenomenological broadening $\Gamma/t=0.01$. In both panels, $\sigma_0=e^2/(4\hbar)$ and $g/t=1$.
}
\label{fig:Conductivity-cyclic-sigma-effective-num}
\end{figure*}

In this section, we calculate the optical conductivity for the cyclic $AB-BC-CA$ stacking. To elucidate the role of the band-crossing points, we compare the interband conductivity for the effective model with that obtained in the tight-binding one in Fig.~\ref{fig:Conductivity-cyclic-sigma-effective-num}. Because there is dependence only on $|\mu-g|$ in the effective model, we show the results for $\mu<g$. As one can see, while the conductivities in both models show plateaus with similar onsets and offsets, there are qualitative differences. In particular, there is no particle-hole symmetry with respect to the band-crossing point in the tight-binding model, which is reflected in the different magnitudes of the conductivity plateaus; see Appendix~\ref{sec:App-Conductivity-cyclic} for the detailed discussion of the results in the effective model. The features at $\hbar\Omega/g \gtrsim 0.3$ are affected by the details of the energy spectrum away from the crossing points.

\begin{figure*}[t]
\centering
\subfigure[]{\includegraphics[width=0.45\textwidth]{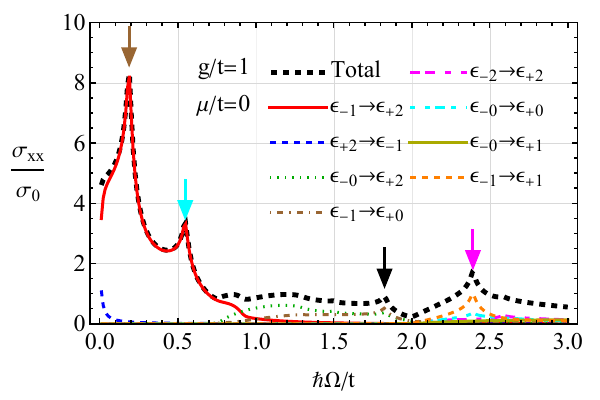}}
\hspace{0.01\textwidth}
\subfigure[]{\includegraphics[width=0.45\textwidth]{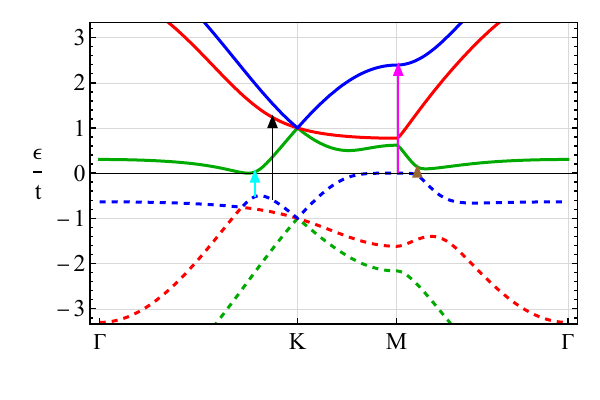}}
\caption{
Panel (a): The interband optical conductivity for the tight-binding model corresponding to the cyclic $AB-BC-CA$ stacking at $\mu=0$. The contributions due to different bands are marked by different lines. Vertical arrows mark the peaks of the optical conductivity; the corresponding transitions are shown in panel (b). We used the phenomenological broadening $\Gamma/t = 0.01$.
Panel (b): The energy spectrum of the tight-binding Hamiltonian along the $\Gamma-\mbox{K}-\mbox{M}-\Gamma$ line in the Brillouin zone. The vertical arrows show the transitions contributing to the peaks in the optical conductivity at $\mu=0$.
In both panels, $\sigma_0=e^2/(4\hbar)$ and $g/t=1$.
}
\label{fig:Conductivity-cyclic-TB-full}
\end{figure*}

In the case of Fermi energies away from the crossing points, we resort to the tight-binding model. The contributions from each of the transitions in the optical conductivity are shown in Fig.~\ref{fig:Conductivity-cyclic-TB-full}(a). The nontrivial band structure in the cyclic stacking leads to a set of noticeable features that are not captured by the effective model. Unlike the case of the aligned $AA-BB-CC$ stacking, the transitions between all types of bands are possible as long as they are not Pauli-blocked. The most prominent peaks in the optical conductivity can be explained by transitions between the extrema of the filled bands (dashed blue line) and the empty bands (solid lines) in Fig.~\ref{fig:Conductivity-cyclic-TB-full}(b).

\begin{figure*}[t]
\centering
\subfigure[]{\includegraphics[width=0.45\textwidth]{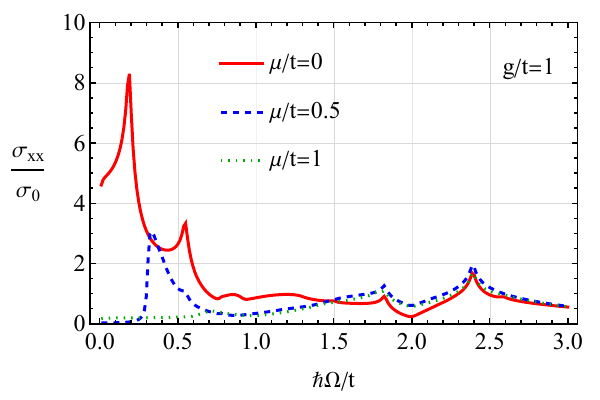}}
\hspace{0.01\textwidth}
\subfigure[]{\includegraphics[width=0.45\textwidth]{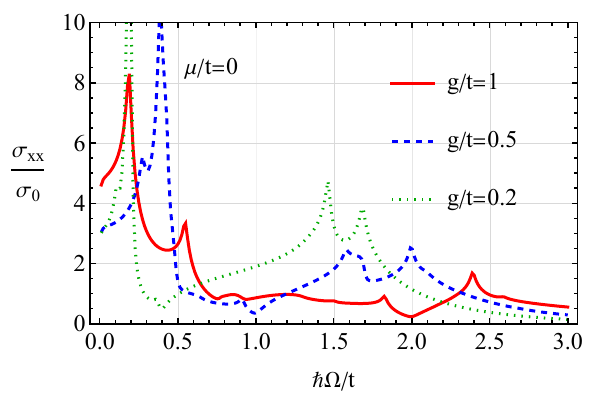}}
\caption{
The interband optical conductivity for the tight-binding model of the cyclic $AB-BC-CA$ stacking for a few values of $\mu$ at $g/t=1$ (panel (a)) and a few values of $g$ at $\mu/t=0$ (panel (b)). In both panels, $\sigma_0=e^2/(4\hbar)$ and we used the phenomenological broadening $\Gamma/t = 0.01$.
}
\label{fig:Conductivity-cyclic-TB-contributions}
\end{figure*}

The optical conductivity at several values of $g/t$ and $\mu/t$ is shown in Figs.~\ref{fig:Conductivity-cyclic-TB-contributions}(a) and \ref{fig:Conductivity-cyclic-TB-contributions}(b), respectively. As expected, the rise of the Fermi energy blocks several transitions leading to the disappearance of the low-frequency peaks while leaving the high-frequency ones intact. The dependence on the coupling constant is nonmonotonic for certain features (low-frequency peaks); the other may be shifted to lower frequencies (e.g., the peaks at $\hbar\Omega/t \gtrsim 1$).

\section{Summary}
\label{sec:Summary}

In this work, we investigated the optical conductivity of bilayer dice (or $\mathcal{T}_3$) lattices introduced in Ref.~\cite{SOG:part1-2023}. A bilayer dice lattice realizes four commensurate stacking: aligned $AA-BB-CC$, hub-aligned $AB-BA-CC$, mixed $AA-BC-CB$, and cyclic $AB-BC-CA$. Each of these stackings has a different energy spectrum and, as a result, distinct interband optical conductivity and activation behavior. To make an analytical advance, we employed effective models valid in the vicinity of the band-crossing points. The results for the tight-binding models are also discussed. The effective models are able to capture the features of the optical conductivity related to the band-crossing $K$ and $K^{\prime}$ points. However, in general, they do not saturate the optical conductivity for all considered stackings.

The optical conductivity for the aligned $AA-BB-CC$ stacking is similar to that of single-layer graphene with, however, a different activation behavior; see Eq.~(\ref{Conductivity-aligned-sigma-xx-fin}). In this case, only the transitions involving the flat band are allowed; see Sec.~\ref{sec:Conductivity-aligned}. There is a good agreement between the effective and tight-binding models at small frequencies signaling that the vicinity of the crossing points provides the main contribution to the optical conductivity. The contributions of the states in the vicinity of the $M$ point of the Brillouin zone become pronounced for larger frequencies or Fermi energies away from the band-crossing points leading to discrepancies between the effective and tight-binding models; see Fig.~\ref{fig:Conductivity-aligned-TB-full}.

In the case of the hub-aligned $AB-BA-CC$ stacking, the band-crossing point plays an important albeit not the dominant role. The corresponding effective model relies only on the transitions involving low- and high-energy dispersive bands but omits the intermediate band; it is well-described by a two-band particle-hole-asymmetric semi-Dirac model. The effective model is able to describe the activation behavior but does not reproduce the shape of the optical conductivity profile well; see Fig.~\ref{fig:Conductivity-hub-sigma-xx-yy-effective-TB}. This discrepancy is explained by contributions from both band-crossing points and other parts of the energy spectrum away from the points. The results for the broader range of Fermi energies and coupling constants reveal a rich structure with a few peaks that can be attributed to the extrema near the $M$ point of the Brillouin zone. The corresponding dependencies are shown in Figs.~\ref{fig:Conductivity-hub-TB-full} and \ref{fig:Conductivity-hub-TB-mu-g}, and can be used to identify the local extrema in the dispersion relation that are responsible for the peaks.

The band-crossing point plays an even less profound role in the mixed $AA-BC-CB$ stacking; see Sec.~\ref{sec:Conductivity-mixed}. This is evident from comparing the interband conductivity obtained in the effective and tight-binding models in Fig.~\ref{fig:Conductivity-mixed-sigma-xx-yy-effective-TB}. In the former, the intermediate band also plays no role allowing us to use a two-band model corresponding to tilted Dirac fermions. Other features in the interband conductivity originate from the transitions between the parts of the energy spectrum away from the band-crossing points; see Fig.~\ref{fig:Conductivity-mixed-TB-full}. The peaks of the optical conductivity can be identified with the transitions between local extrema including those in the vicinity of the $M$ point. As is clear from Fig.~\ref{fig:Conductivity-mixed-TB-full}, all bands may contribute to the optical conductivity leading to an intricate profile with several peaks; see also Fig.~\ref{fig:Conductivity-mixed-TB-contributions}.

Finally, the interband optical conductivity for the cyclic $AB-BC-CA$ stacking reveals a plateau-like feature determined by the interplay of the transitions between occupied and empty states; see Sec.~\ref{sec:Conductivity-cyclic}. The effective model correctly captures the onset and offset of the plateaus but misses particle-hole asymmetry with respect to the band-crossing point; see Fig.~\ref{fig:Conductivity-cyclic-sigma-effective-num}. As in the case of the hub-aligned and mixed stackings, there are no restrictions on the transitions between the bands as long as they are permitted by the Pauli principle. This is revealed in several peaks in the optical conductivity originating from various interband transitions; see Figs.~\ref{fig:Conductivity-cyclic-TB-full} and \ref{fig:Conductivity-cyclic-TB-contributions}.

Thus, we found that optical conductivity provides an effective way to probe the nontrivial dispersion relation, quantify the inter-layer coupling, and distinguish between various commensurate stackings in bilayer dice models. In particular, the optical response of a band-crossing point is manifested in distinct steplike features with a different activation behavior for each of the stacking. For larger frequencies $\hbar \Omega \gtrsim g$, the transitions involving both band-crossing points as well as other parts of the energy spectrum become relevant.

It is noticeable also that the commensurate stackings of the dice lattices (with the exception of the aligned $AA-BB-CC$ stacking) generically lack forbidden transitions leading to several peaks in the optical conductivity. Furthermore, even for large coupling constants and the Fermi energy close to the band-crossing points, the vicinity of the band-crossing may not saturate the optical conductivity leading to noticeable discrepancies between the effective and tight-binding models. Therefore, the studies of optical conductivity may be used to glean information about the structure of the energy bands and the presence of local extrema there. Among the latter, we notice the states near the $M$ point of the Brillouin zone.

In the present work, we focused mostly on the interband transitions and described the effects of disorder phenomenologically by introducing an energy-independent broadening in the tight-binding models. We leave a more detailed investigation of disorder effects for future studies~\footnote{Notice that disorder effects in flat-band systems require special attention, see, e.g., Ref.~\cite{Louvet-Carpentier:2015,Gorbar-Oriekhov:2018,Bouzerar-Mayou:2020,Wang-Ting:2020,Wang-Peeters:2021,Huhtinen-Torma:2022} for recent studies.}. Another perspective direction will be to investigate higher-order optical responses including the second harmonic generation and rectification.

\begin{acknowledgments}
P.O.S. acknowledges support through the Yale Prize Postdoctoral Fellowship in Condensed Matter Theory. D.O.O. acknowledges the support from the Netherlands Organization for Scientific Research (NWO/OCW) and from the European Research Council (ERC) under the European Union's Horizon 2020 research and innovation program.
\end{acknowledgments}

\appendix

\section{Non-abbreviated effective models}
\label{sec:App-Models-full}

For the sake of completeness, we present the effective models retaining all terms up to the second order in $\hbar v_F k/g$ and the first order in $\hbar v_F ak^2/g$ for the hub-aligned $AB-BA-CC$ and mixed $AA-BC-CB$ stackings~\cite{SOG:part1-2023}.

The corresponding effective Hamiltonian for the hub-aligned stacking reads
\begin{widetext}
\begin{equation}
\label{App-Model-effective-hub-Heff}
H_{\rm eff}^{\rm (h)} = g \mathds{1}_3 + \frac{\hbar v_F}{\sqrt{2}} k_x \left(
                                                            \begin{array}{ccc}
                                                              0 & 1 & 0 \\
                                                              1 & 0 & 1 \\
                                                              0 & 1 & 0 \\
                                                            \end{array}
                                                          \right) +\left(\frac{\hbar v_F}{\sqrt{2}} \right)^2 \frac{k_y^2}{2g}\left(
                                                                     \begin{array}{ccc}
                                                                       1 & 0 & -1 \\
                                                                       0 & 2 & 0 \\
                                                                       -1 & 0 & 1 \\
                                                                     \end{array}
                                                                   \right)
                                                          + \frac{\hbar v_F}{\sqrt{2}} \frac{a}{4} \left(k_y^2 -k_x^2\right) \left(
                                                            \begin{array}{ccc}
                                                              0 & 1 & 0 \\
                                                              1 & 0 & 1 \\
                                                              0 & 1 & 0 \\
                                                            \end{array}
                                                          \right).
\end{equation}
\end{widetext}
Its abbreviated version is given in Eq.~(\ref{Model-effective-hub-Heff}).

In the case of the mixed stacking, the effective Hamiltonian is
\begin{widetext}
\begin{eqnarray}
\label{App-Model-effective-mixed-Heff}
H_{\rm eff}^{\rm (m)} &=& g \mathds{1}_3 + \frac{\hbar v_F}{2\sqrt{2}} \left(
                                                            \begin{array}{ccc}
                                                              0 & 2k_x & k_{-} \\
                                                              2k_x & 0 & k_{-} \\
                                                              k_{+} & k_{+} & 0 \\
                                                            \end{array}
                                                          \right) + \left(\frac{\hbar v_F}{4} \right)^2 \frac{1}{g}\left(
                                        \begin{array}{ccc}
                                          k_x^2 +5k_y^2 & 0 & 0 \\
                                          0 & k_x^2 +5k_y^2 & 0 \\
                                          0 & 0 & 2k^2 \\
                                        \end{array}
                                      \right) \nonumber\\
&-& \left(\frac{\hbar v_F}{4} \right)^2 \frac{1}{g}\left(
                                        \begin{array}{ccc}
                                          0 & k^2 & 2ik_y k_{-} \\
                                          k^2 & 0 & 2ik_y k_{-} \\
                                          -2ik_y k_{+} & -2ik_y k_{+} & 0 \\
                                        \end{array}
                                      \right)
- \frac{\hbar v_F a}{8\sqrt{2}} \left(
                                        \begin{array}{ccc}
                                          0 & 2(k_x^2-k_y^2) & k_{+}^2 \\
                                          2(k_x^2-k_y^2) & 0 & k_{+}^2 \\
                                          k_{-}^2 & k_{-}^2 & 0 \\
                                        \end{array}
                                      \right).
\end{eqnarray}
\end{widetext}
The corresponding abbreviated version is given in Eq.~(\ref{Model-effective-mixed-Heff}).

As one can see, even the effective models for the hub-aligned $AB-BA-CC$ and mixed $AA-BC-CB$ stackings are rather cumbersome and inconvenient for analytical analysis. Nevertheless, to show that the abbreviated models capture the main features of the optical conductivity, we compare the conductivity for non-abbreviated and abbreviated models in Figs.~\ref{fig:Conductivity-hub-sigma-xx-yy} and \ref{fig:Conductivity-mixed-sigma-xx-yy}.

\section{Calculation of optical conductivity}
\label{sec:App-Conductivity}

In this appendix, we provide the details of the calculation of the optical conductivity in the effective models; see Sec.~\ref{sec:Conductivity} for the definitions and comparison of the final results with tight-binding models. We use the Kubo linear response approach discussed in Sec.~\ref{sec:Conductivity-Kubo}.

\subsection{Aligned \texorpdfstring{$AA-BB-CC$}{AA-BB-CC} stacking}
\label{sec:App-Conductivity-aligned}

The effective Hamiltonian for the aligned $AA-BB-CC$ stacking is given in Eq.~(\ref{Model-effective-aligned-Heff}). The velocity matrix $v_n =  \partial_{k_n} H(\mathbf{k})/\hbar$ reads as
\begin{equation}
\label{App-Conductivity-aligned-v}
\mathbf{v} = \frac{v_F}{\sqrt{2}} \mathbf{S},
\end{equation}
where the (psudo)spin-1 matrices $S_x$ and $S_y$ are given in Eq.~(\ref{Model-S-def}).

By using Eqs.~(\ref{Model-effective-aligned-Heff}), (\ref{Conductivity-G-def}), and (\ref{Conductivity-A-def}), we derive the following traces in Eq.~(\ref{Conductivity-Kubo-sigma-nn}):
\begin{widetext}
\begin{eqnarray}
\label{App-Conductivity-aligned-vxAvxA}
&&\mbox{Tr}{\left(v_x A(\omega; \mathbf{k}) v_x A(\omega-\Omega; \mathbf{k}) \right)} =
2v_F^2 F(\omega) F(\omega -\Omega) \nonumber\\
&&\quad\times \left\{\left(\hbar \omega -\hbar \Omega - g\right)^2\left(\hbar \omega - g\right)^2 -\frac{(\hbar v_F k)^2}{4}\left[(\hbar\Omega)^2 - \left(\hbar\Omega -2\hbar \omega - 2g\right)^2 \cos{(2\varphi)} \right] \right\},\\
\label{App-Conductivity-aligned-vyAvyA}
&&\mbox{Tr}{\left(v_y A(\omega; \mathbf{k}) v_y A(\omega-\Omega; \mathbf{k}) \right)} =
2v_F^2 F(\omega) F(\omega -\Omega) \nonumber\\
&&\quad\times \left\{\left(\hbar \omega -\hbar \Omega - g\right)^2\left(\hbar \omega - g\right)^2 -\frac{(\hbar v_F k)^2}{4}\left[(\hbar\Omega)^2 + \left(\hbar\Omega -2\hbar \omega + 2g\right)^2 \cos{(2\varphi)} \right] \right\}.
\end{eqnarray}
\end{widetext}
Here,
\begin{equation}
\label{App-Conductivity-aligned-F-def}
F(\omega)= \sum_{n=0}^{2} \frac{\delta(\hbar\omega -\epsilon_{n})}{\Pi_{m=0}^{2,\prime}(\epsilon_n -\epsilon_{m})},
\end{equation}
where the product $\Pi_{m=0}^{2, \prime} (\epsilon_{n}-\epsilon_{m})$ excludes $\epsilon_{n} = \epsilon_{m}$. The energy spectrum is $\epsilon_{0} = g$, $\epsilon_{1} =  g + \hbar v_F k$, and $\epsilon_{2} = g - \hbar v_F k$.

By substituting Eq.~(\ref{App-Conductivity-aligned-vxAvxA}) into Eq.~(\ref{Conductivity-Kubo-sigma-nn}) and calculating integrals over $\varphi$ and $\omega$, we obtain
\begin{widetext}
\begin{eqnarray}
\label{App-Conductivity-aligned-sigma-xx}
\mbox{Re}{\left\{\sigma_{xx}\right\}} &=& -\frac{e^2}{\hbar \Omega} \int_0^{\infty}
\frac{dk}{v_F^2 k^3}
\Bigg\{
-\delta(\Omega) \frac{(v_Fk \Omega)^2}{4} \left[f^{\rm eq}(g) -f^{\rm eq}(g-\hbar\Omega)\right] \nonumber\\
&+&\delta(\Omega) \frac{1}{4}\left[\left(\Omega-v_Fk\right)^2(v_F k)^2 -\frac{(v_Fk \Omega)^2}{4}\right] \left[f^{\rm eq}(g+\hbar v_Fk) -f^{\rm eq}(g+\hbar v_Fk-\hbar\Omega)\right] \nonumber\\
&+&\delta(\Omega) \frac{1}{4}\left[\left(\Omega+v_Fk\right)^2(v_F k)^2 -\frac{(v_Fk \Omega)^2}{4}\right] \left[f^{\rm eq}(g- \hbar v_Fk) -f^{\rm eq}(g-\hbar v_Fk-\hbar\Omega)\right] \nonumber\\
&+&\left[\delta(\Omega-v_Fk) +\delta(\Omega+v_Fk)\right] \frac{(v_F k\Omega)^2}{8} \left[f^{\rm eq}(g+\hbar\Omega) -f^{\rm eq}(g-\hbar \Omega)\right] \nonumber\\
&+& \delta(\Omega+2v_Fk) \frac{1}{4}\left[\left(\Omega+v_Fk\right)^2(v_Fk)^2 -\frac{(v_F k\Omega)^2}{4}\right] \left[f^{\rm eq}(g-\hbar v_F k) -f^{\rm eq}(g-\hbar v_F k -\hbar \Omega)\right] \nonumber\\
&+& \delta(\Omega-2v_Fk) \frac{1}{4}\left[\left(\Omega-v_Fk\right)^2(v_Fk)^2 -\frac{(v_F k\Omega)^2}{4}\right] \left[f^{\rm eq}(g+\hbar v_F k) -f^{\rm eq}(g+\hbar v_F k -\hbar \Omega)\right]
\Bigg\}.
\end{eqnarray}
\end{widetext}
In the case of interband conductivity, only the terms with $\delta(\Omega\pm v_Fk)$ contribute; the prefactor at $\delta(\Omega\pm 2v_Fk)$ vanishes after integrating over $k$. This means that there are no direct transitions between the dispersive bands, i.e., $\epsilon_{1,2} =g \pm \hbar v_F k$. The final result for $\mbox{Re}{\left\{\sigma_{xx}\right\}}$ is given in Eq.~(\ref{Conductivity-aligned-sigma-xx-fin}). Finally, by substituting Eq.~(\ref{App-Conductivity-aligned-vyAvyA}) into Eq.~(\ref{Conductivity-Kubo-sigma-nn}) and integrating over $\varphi$, it is straightforward to show that $\sigma_{xx}=\sigma_{yy}$; the absence of the Hall components $\sigma_{xy}=\sigma_{yx}=0$ follows from the time-reversal symmetry.

\subsection{Hub-aligned \texorpdfstring{$AB-BA-CC$}{AB-BA-CC} stacking}
\label{sec:App-Conductivity-hub}

In this section, we provide the details of calculations of the conductivity for the effective model of the hub-aligned $AB-BA-CC$ stacking. We use the abbreviated effective Hamiltonian given in Eq.~(\ref{Model-effective-hub-Heff}) and focus on the contribution of a single $K$ point. We have the following components of the velocity matrix:
\begin{equation}
\label{App-Conductivity-hub-v}
v_x = \frac{v_F}{\sqrt{2}} \left(
                                 \begin{array}{ccc}
                                   0 & 1 & 0 \\
                                   1 & 0 & 1 \\
                                   0 & 1 & 0 \\
                                 \end{array}
                               \right) \quad \mbox{and} \quad
v_y = \frac{\hbar v_F^2 k_y}{2g} \left(
                                 \begin{array}{ccc}
                                   1 & 0 & -1 \\
                                   0 & 2 & 0 \\
                                   -1 & 0 & 1 \\
                                 \end{array}
                               \right).
\end{equation}

The traces in Eq.~(\ref{Conductivity-Kubo-sigma-nn}) are
\begin{widetext}
\begin{eqnarray}
\label{App-Conductivity-hub-vxAvxA}
&&\mbox{Tr}{\left(v_x A(\omega; \mathbf{k}) v_x A(\omega-\Omega; \mathbf{k}) \right)} =
F(\omega) F(\omega -\Omega) \frac{v_F^2}{8g^3} \Big\{(\hbar v_F k)^2 \left[4g \cos^2{(\varphi)} +(2g - 2\hbar \omega + \hbar \Omega) \sin^2{(\varphi)}\right] \nonumber\\
&&\quad +4g(g - \hbar \omega)(g - \hbar \omega + \hbar \Omega)\Big\} \left\{2g(g - \hbar \omega + \hbar \Omega) +\left[\hbar v_F k \sin{(\varphi)}\right]^2\right\} \left\{2g(g - \hbar \omega) +\left[\hbar v_F k \sin{(\varphi)}\right]^2\right\}, \\
\label{App-Conductivity-hub-vyAvyA}
&&\mbox{Tr}{\left(v_y A(\omega; \mathbf{k}) v_y A(\omega-\Omega; \mathbf{k}) \right)} =
F(\omega) F(\omega -\Omega)
\frac{v_F^2 \left[\hbar v_F k \sin{(\varphi)}\right]^2}{8g^4} \Bigg\{8g^2 \left[\hbar v_F k \cos{(\varphi)}\right]^4 \nonumber\\
&&\quad+(g - \hbar \omega) (g - \hbar \omega + \hbar \Omega) \left[(\hbar v_F k)^2 +4g(g - \hbar \omega + \hbar \Omega) -(\hbar v_F k)^2 \cos{(2\varphi)}\right] \nonumber\\
&&\quad \times \left[(\hbar v_F k)^2 +4g(g- \hbar \omega) -(\hbar v_F k)^2 \cos{(2\varphi)}\right] -2g\left[\hbar v_F k \cos{(\varphi)}\right]^2 \Big\{2(g - \hbar \omega)\left[4g (g - \hbar \omega) +(\hbar v_F k)^2\right]\nonumber\\
&&\quad +\hbar \Omega \left[8g (g - \hbar \omega) +(\hbar v_F k)^2\right]
+4g (\hbar \Omega)^2 -(\hbar v_F k)^2 (2g - 2\hbar \omega + \hbar \Omega) \cos{(2\varphi)}
\Big\}
\Bigg\}.
\end{eqnarray}
\end{widetext}
Here, we used Eqs.~(\ref{Model-effective-hub-Heff}), (\ref{Conductivity-G-def}), and (\ref{Conductivity-A-def}); $F(\omega)$ is defined in Eq.~(\ref{App-Conductivity-aligned-F-def}) and the energy dispersion $\epsilon_{n=0,1,2}$ is given in Eqs.~(\ref{Model-effective-hub-eps-0})--(\ref{Model-effective-hub-eps-2}). Notice that the traces in Eqs.~(\ref{App-Conductivity-hub-vxAvxA}) and (\ref{App-Conductivity-hub-vyAvyA}) are noticeably different. As we show in Appendix~\ref{sec:App-Conductivity-SD}, this leads to an anisotropic conductivity $\sigma_{xx} \neq \sigma_{yy}$ at a given band-crossing point; the isotropy is restored after averaging over all crossing points in the Brillouin zone.

The integrals over $\omega$ in Eqs.~(\ref{App-Conductivity-hub-vxAvxA}) and (\ref{App-Conductivity-hub-vyAvyA}) have the following form:
\begin{widetext}
\begin{eqnarray}
\label{App-Conductivity-hub-FF-int}
&&\int d \omega \, F(\omega) F(\omega -\Omega) B(\omega) = \sum_{n_1,n_2=0}^{2} \int d \omega \frac{\delta(\hbar \omega -\epsilon_{n_1}) \delta(\hbar\omega -\hbar\Omega- \epsilon_{n_2})}{\Pi_{m_1=0}^{2, \prime} (\epsilon_{n_1}-\epsilon_{m_1}) \Pi_{m_2=0}^{2, \prime} (\epsilon_{n_2}-\epsilon_{m_2})} B(\omega) \nonumber\\
&&\quad =\sum_{n_1=n_2}^{2} \frac{\delta(\hbar\Omega) B(\epsilon_{n_1})}{\Pi_{m_1=0}^{2, \prime} (\epsilon_{n_1}-\epsilon_{m_1}) \Pi_{m_2=0}^{2, \prime} (\epsilon_{n_1}-\epsilon_{m_2})}
+\!\sum_{n_1\neq n_2}^{2} \frac{\delta(\epsilon_{n_1} -\epsilon_{n_2} -\hbar\Omega) B(\epsilon_{n_1})}{\Pi_{m_1=0}^{2, \prime} (\epsilon_{n_1}-\epsilon_{m_1}) \Pi_{m_2=0}^{2, \prime} (\epsilon_{n_2} -\epsilon_{m_2})}.
\end{eqnarray}
\end{widetext}
The terms with $\delta(\hbar\Omega)$ and $\delta(\epsilon_{n_1} -\epsilon_{n_2} -\hbar\Omega)$ correspond to the intra- and interband transitions, respectively. In the case of effective models, we focus on the interband transitions and omit the intra-band terms. This allows us to integrate over $k$ in the conductivity analytically. The corresponding results are cumbersome; therefore, we do not present them here. We notice, however, that in the resulting expression, one only needs to integrate over $\varphi$.

\subsection{Particle-hole asymmetric semi-Dirac model}
\label{sec:App-Conductivity-SD}

In this Section, we calculate the optical conductivity for a particle-hole asymmetric 2D semi-Dirac model given by the following Hamiltonian:
\begin{equation}
\label{Conductivity-hub-SD-H-def}
H_{\rm sD} = \left[g +\frac{(\hbar v_F k_y)^2}{4g}\right] \mathds{1}_2 + \hbar v_F k_x \sigma_x +\frac{(\hbar v_F k_y)^2}{4g} \sigma_y.
\end{equation}
This model is sufficient for describing the interband optical conductivity for the hub-aligned $AB-BA-CC$ stacking because only the $\epsilon_{n=1}$ and $\epsilon_{n=2}$ branches contribute to the interband conductivity in the effective model; see Eqs.~(\ref{Model-effective-mixed-eps-0})--(\ref{Model-effective-mixed-eps-2}) for the dispersion relation. This property can be verified by direct calculation using the results of Appendix~\ref{sec:App-Conductivity-hub}. Such selection rules are qualitatively different from the case of the aligned $AA-BB-CC$ stacking discussed in Sec.~\ref{sec:Conductivity-aligned} where only transitions involving the $\epsilon_{n=0}$ band are allowed. We notice also that the optical conductivity of a particle-hole symmetric version of Hamiltonian (\ref{Conductivity-hub-SD-H-def}) (i.e., without the first term) was calculated in Ref.~\cite{Carbotte-Nicol:2019}.

The Green function for Hamiltonian (\ref{Conductivity-hub-SD-H-def}) reads as
\begin{widetext}
\begin{equation}
\label{App-Conductivity-SD-G-def}
G(\omega;\mathbf{k}) = \frac{1}{\mathcal{D}} \left[\left(\hbar \omega-g-\frac{(\hbar v_F k_y)^2}{4g}\right)\mathds{1}_2 +\hbar v_F k_x \sigma_x + \frac{(\hbar v_F k_y)^2}{4g} \sigma_y\right],
\end{equation}
\end{widetext}
where the denominator is
\begin{eqnarray}
\label{App-Conductivity-SD-D-def}
\mathcal{D} &=& \prod_{\eta=\pm}\left(\hbar\omega - \epsilon_{\eta}\right) = \left[\hbar\omega -g -\frac{(\hbar v_F k_y)^2}{4g}\right]^2 \nonumber\\
&-&\left\{(\hbar v_F k_x)^2 +\left[\frac{(\hbar v_F k_y)^2}{4g}\right]^2\right\}.
\end{eqnarray}
For the sake of convenience, here and henceforth we use $\epsilon_{\pm}\equiv \epsilon_{1,2}$; see Eqs.~(\ref{Model-effective-mixed-eps-0})--(\ref{Model-effective-mixed-eps-2}) for the definition of $\epsilon_{n=1,2}$.

The spectral function (\ref{Conductivity-A-def}) is
\begin{widetext}
\begin{equation}
\label{App-Conductivity-SD-A-def}
A(\omega;\mathbf{k}) =\left[\left(\hbar \omega -g -\frac{(\hbar v_F k_y)^2}{4g}\right)\mathds{1}_2 +\hbar v_F k_x \sigma_x + \frac{(\hbar v_F k_y)^2}{4g} \sigma_y\right] \sum_{\eta=\pm} \frac{\delta(\hbar \omega - \epsilon_{\eta})}{\epsilon_{\eta}-\epsilon_{-\eta}}.
\end{equation}
\end{widetext}

The traces in the conductivity (\ref{Conductivity-Kubo-sigma-nn}) are
\begin{widetext}
\begin{eqnarray}
\label{App-Conductivity-SD-vxGvxG}
&&\mbox{tr} \left[ v_x A(\omega; \mathbf{k}) v_x A(\omega-\Omega; \mathbf{k})\right] = 2v_F^2 \left\{(\hbar \omega -g) (\hbar\omega -g - \hbar\Omega) +K^2 \cos^2{(\phi)} +\left[\hbar\Omega -2(\hbar\omega-g)\right] K \sin{(\phi)} \right\} \nonumber\\
&&\times \sum_{\eta_1,\eta_2=\pm} \frac{\delta(\hbar \omega - \epsilon_{\eta_1})}{\epsilon_{\eta_1}-\epsilon_{-\eta_1}} \frac{\delta(\hbar \omega - \hbar \Omega +\epsilon_{\eta_2})}{\epsilon_{\eta_2}-\epsilon_{-\eta_2}},\\
\label{App-Conductivity-SD-vyGvyG}
&&\mbox{tr} \left[ v_y A(\omega; \mathbf{k}) v_y A(\omega-\Omega; \mathbf{k})\right] = \frac{4v_F^2}{g} (\hbar\omega -g)(\hbar \omega -g -\hbar\Omega) K \sin{(\phi)} \!\! \sum_{\eta_1,\eta_2=\pm} \!\! \frac{\delta(\hbar \omega - \epsilon_{\eta_1})}{\epsilon_{\eta_1}-\epsilon_{-\eta_1}} \frac{\delta(\hbar \omega - \hbar \Omega +\epsilon_{\eta_2})}{\epsilon_{\eta_2}-\epsilon_{-\eta_2}},
\end{eqnarray}
\end{widetext}
where we introduced the following variables:
\begin{equation}
\label{App-Conductivity-SD-K-phi}
\hbar v_F k_x = K \cos{(\phi)} \quad \mbox{and} \quad \frac{(\hbar v_F k_y)^2}{4g} = K \sin{(\phi)}
\end{equation}
with $0\leq\phi\leq\pi/2$ and $K\geq0$. The corresponding Jacobian is
\begin{equation}
\label{App-Conductivity-SD-Jacobian}
J(K,\phi) = 4\frac{\partial (k_x, k_y)}{\partial (K, \phi)} = \frac{4}{(\hbar v_F)^2} \sqrt{\frac{g K}{\sin{(\phi)}}};
\end{equation}
the additional factor $4$ originates from the integration range $0\leq\phi\leq\pi/2$.

The new variables (\ref{App-Conductivity-SD-K-phi}) allow us to rewrite the energies $\epsilon_{\pm}$ in a simple form
\begin{equation}
\label{App-Conductivity-SD-eps-new}
\epsilon_{\eta} =  g + K \left[\sin{(\phi)} +\eta\right],
\end{equation}
where $\eta=\pm$.

By using Eqs.~(\ref{Conductivity-Kubo-sigma-nn}), (\ref{App-Conductivity-SD-vxGvxG}), and (\ref{App-Conductivity-SD-eps-new}), we obtain the following real part of the conductivity $\mbox{Re}{\left\{\sigma_{xx}(\Omega)\right\}}$:
\begin{widetext}
\begin{eqnarray}
\label{App-Conductivity-SD-sigma-xx}
&&\mbox{Re}{\left\{\sigma_{xx}(\Omega)\right\}} = -\frac{2\hbar v_F^2 \sigma_0}{\Omega} \sum_{\eta_1,\eta_2 =\pm} \int_0^{\pi/2} \frac{d\phi}{2\pi} \int_0^{\infty} dK\, \left[f^{\rm eq}\left(g +K \left[\sin{(\phi)}+\eta_1\right]\right) -f^{\rm eq}\left(g +K \left[\sin{(\phi)}+\eta_1\right] -\hbar \Omega\right)\right]  \nonumber\\
&&\times J(K,\phi) \left\{K \left[\sin{(\phi)}+\eta_1\right] \left\{K \left[\sin{(\phi)}+\eta_1\right] - \hbar \Omega\right\} +K^2 \cos^2{(\phi)}
-K\sin{(\phi)} \left\{2K \left[\sin{(\phi)}+\eta_1\right] -\hbar \Omega\right\} \right\} \nonumber\\
&&\times \frac{\delta\left((\eta_1 - \eta_2)K -\hbar \Omega\right)}{\eta_1\eta_2K^2}.
\end{eqnarray}
\end{widetext}
Here, the case $\eta_1=\eta_2$ corresponds to intra-band transitions and we used $\sigma_0=e^2/(4\hbar)$. We focus only on the interband transitions for which $\eta_1 = -\eta_2 =1$:
\begin{widetext}
\begin{eqnarray}
\label{App-Conductivity-SD-sigma-xx-inter}
&&\mbox{Re}{\left\{\sigma_{xx}^{\rm (inter)}(\Omega)\right\}} =
\frac{2\hbar v_F^2 \sigma_0}{\Omega} \int_0^{\pi/2} \frac{d\phi}{2\pi} \int_0^{\infty} dK\, \left\{f^{\rm eq}\left(g +K\left[\sin{(\phi)}+1\right]\right) -f^{\rm eq}\left(g +K\left[\sin{(\phi)}+1\right] -\hbar \Omega\right)\right\} J(K,\phi) \nonumber\\
&&\times \left\{K\left[\sin{(\phi)}+1\right] \left\{K\left[\sin{(\phi)}+1\right] - \hbar \Omega\right\} +K^2 \cos^2{(\phi)} -K\sin{(\phi)} \left\{2K \left[\sin{(\phi)}+1\right] -\hbar \Omega\right\}\right\} \frac{\delta\left(2K -\hbar \Omega\right)}{K^2} \nonumber\\
&&= \frac{2\hbar v_F^2 \sigma_0}{\Omega}  \int_0^{\pi/2} \frac{d\phi}{2\pi} \left\{f^{\rm eq}\left(g +\frac{\hbar\Omega}{2}\left[\sin{(\phi)}-1\right]\right) -f^{\rm eq}\left(g +\frac{\hbar\Omega}{2}\left[\sin{(\phi)}+1\right]\right)\right\} J\left(\frac{\hbar \Omega}{2},\phi\right) \sin^2{(\phi)} \nonumber\\
&&= 4\sigma_0 \sqrt{\frac{2g}{\hbar\Omega}} \int_0^{\pi/2} \frac{d\phi}{2\pi} \sin^{3/2}{(\phi)} \left\{f^{\rm eq}\left(g +\frac{\hbar\Omega}{2}\left[\sin{(\phi)}-1\right]\right) -f^{\rm eq}\left(g +\frac{\hbar\Omega}{2}\left[\sin{(\phi)}+1\right]\right)\right\},
\end{eqnarray}
\end{widetext}
where we used Eq.~(\ref{App-Conductivity-SD-Jacobian}) in the last line.
The integral over $\phi$ can be taken numerically. It is also possible to calculate it analytically for $T\to0$ in terms of hypergeometric functions; however, the corresponding expressions are too bulky to be presented here.

The real part of the conductivity $\mbox{Re}{\left\{\sigma_{yy}(\Omega)\right\}}$ is
\begin{widetext}
\begin{eqnarray}
\label{App-Conductivity-SD-sigma-yy}
\mbox{Re}{\left\{\sigma_{yy}(\Omega)\right\}} &=& -\frac{4\hbar v_F^2 \sigma_0}{g\Omega}\sum_{\eta_1,\eta_2 =\pm} \int_0^{\pi/2} \frac{d\phi}{2\pi} \int_0^{\infty} dK\, \left\{f^{\rm eq}\left(g+ K \left[\sin{(\phi)} +\eta_1\right]\right) -f^{\rm eq}\left(g+ K \left[\sin{(\phi)} +\eta_1\right] -\hbar \Omega\right)\right\}  \nonumber\\
&\times& J(K,\phi) \sin{(\phi)}\left[\sin{(\phi)} +\eta_1\right] \left\{K \left[\sin{(\phi)} +\eta_1\right] - \hbar \Omega\right\} \frac{\delta\left((\eta_1 - \eta_2)K -\hbar \Omega\right)}{\eta_1\eta_2}.
\end{eqnarray}
\end{widetext}
As with the $xx$ component, we focus only on the interband part with $\eta_1 = -\eta_2 =1$:
\begin{widetext}
\begin{eqnarray}
\label{App-Conductivity-SD-sigma-yy-inter}
&&\mbox{Re}{\left\{\sigma_{yy}^{\rm (inter)}(\Omega)\right\}} =
\frac{4\hbar \sigma_0 v_F^2}{g \Omega} \int_0^{\pi/2} \frac{d\phi}{2\pi} \int_0^{\infty} dK\, \left\{f^{\rm eq}\left(g +K \left[\sin{(\phi)} +1\right]\right) -f^{\rm eq}\left(g +K \left[\sin{(\phi)} +1\right] -\hbar \Omega\right)\right\}  \nonumber\\
&&\times J(K,\phi) \sin{(\phi)} \left[\sin{(\phi)} +1\right] \left\{K \left[\sin{(\phi)} +1\right] -\hbar \Omega \right\} \delta\left(2K -\hbar \Omega\right) \nonumber\\
&&= \frac{(\hbar v_F)^2 \sigma_0}{g} \!\int_0^{\pi/2}\! \frac{d\phi}{2\pi} \left\{\! f^{\rm eq}\left(g +\frac{\hbar\Omega}{2}\left[\sin{(\phi)} -1\right]\right) -f^{\rm eq}\left(g +\frac{\hbar\Omega}{2}\left[\sin{(\phi)} +1\right]\right)\!\right\}\! J\left(\frac{\hbar\Omega}{2},\phi\!\right) \cos^2{(\phi)}\sin{(\phi)} \nonumber\\
&&= 2\sigma_0 \sqrt{\frac{2\hbar \Omega}{g}} \int_0^{\pi/2} \frac{d\phi}{2\pi} \cos^2{(\phi)} \sin^{1/2}{(\phi)} \left\{f^{\rm eq}\left(g +\frac{\hbar\Omega}{2} \left[\sin{(\phi)} -1\right]\right) -f^{\rm eq}\left(g +\frac{\hbar\Omega}{2}\left[\sin{(\phi)} +1\right] \right)\right\},
\end{eqnarray}
\end{widetext}
where we used Eq.~(\ref{App-Conductivity-SD-Jacobian}) in the last line of Eq.~(\ref{App-Conductivity-SD-sigma-yy-inter}).

The conductivity components scale as $\mbox{Re}{\left\{\sigma_{xx}(\Omega)\right\}} \sim (\hbar \Omega/g)^{-1/2}$ and $\mbox{Re}{\left\{\sigma_{yy}(\Omega)\right\}} \sim (\hbar \Omega/g)^{1/2}$ for $\hbar\Omega/g \gg1$ and $\mu>g$ making their product frequency-independent. Due to the absence of the particle-hole symmetry, this is no longer the case for $\mu<g$. Notice that while the regime $\hbar\Omega/g \gg1$ is beyond the applicability of the effective models for the bilayer dice lattice, the corresponding scalings may be useful for other realizations of particle-hole-asymmetric semi-Dirac models.

\begin{figure*}[!ht]
\centering
\subfigure[]{\includegraphics[width=0.45\textwidth]{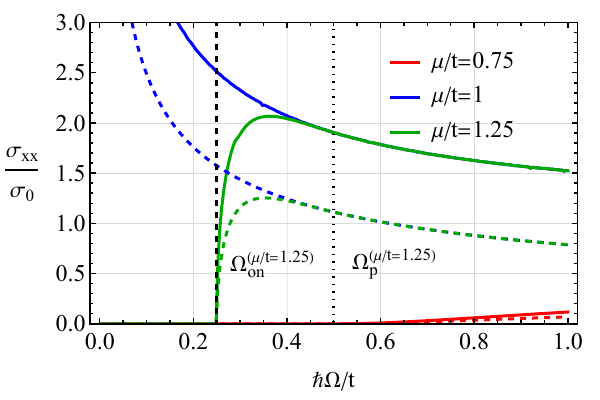}}
\hspace{0.01\textwidth}
\subfigure[]{\includegraphics[width=0.45\textwidth]{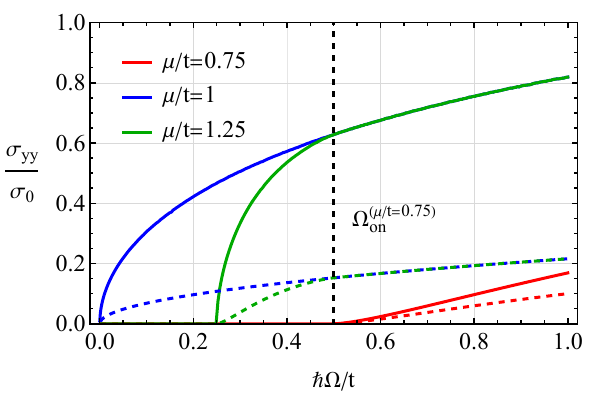}}
\caption{
The dependence of the interband conductivity components $\RE{\sigma_{xx}}$ (panel (a)) and $\RE{\sigma_{yy}}$ (panel (b)) normalized to $\sigma_0 = e^2/(4\hbar)$ for the hub-aligned $AB-BA-CC$ stacking on $\hbar\Omega/t$ for a few values of $\mu/t$. Solid and dashed lines correspond to the non-abbreviated and abbreviated effective models, respectively; see Eqs.~(\ref{App-Model-effective-hub-Heff}) and (\ref{Conductivity-hub-SD-H-def}) for the definitions of the models. In all panels, we used $g/t=1$ and set $T\to 0$. Vertical dashed and dotted lines correspond to the onset and plateau frequencies discussed in the text.
}
\label{fig:Conductivity-hub-sigma-xx-yy}
\end{figure*}

We show $\mbox{Re}{\left\{\sigma_{xx}(\Omega)\right\}}$ and $\mbox{Re}{\left\{\sigma_{yy}(\Omega)\right\}}$ in Fig.~\ref{fig:Conductivity-hub-sigma-xx-yy}. We used the non-abbreviated effective model retaining all terms up to the second order in momentum (solid lines), see Eq.~(\ref{App-Model-effective-hub-Heff}), and the abbreviated effective model (dashed lines), see Eq.~(\ref{Model-effective-hub-Heff}) or (\ref{Conductivity-hub-SD-H-def}). While there are quantitative differences, the models agree well in predicting the activation behavior and the overall shape of the conductivity profile. Therefore, our use of the simplified model in Eq.~(\ref{Conductivity-hub-SD-H-def}) is justified.

The activation behavior of the interband part of the optical conductivity observed in Fig.~\ref{fig:Conductivity-hub-sigma-xx-yy} can be straightforwardly deduced from the energy dispersion.

The onset frequency $\Omega_{\rm on}$ for the completely filled lower band is determined by the minimal distance between empty states at the $\epsilon_{n=1}$ branch and filled states at the $\epsilon_{n=2}$ branch; for $\mu>g$ such a minimal distance is realized at $k_x=0$. There is also a saturation (plateau) frequency $\Omega_{\rm p}$ for which the whole $\epsilon_{n=2}$ branch can contribute to the transitions. Both onset $\Omega_{\rm on}\approx |\mu-g|$ and plateau $\Omega_{\rm p}\approx 2|\mu-g|$ frequencies are in good agreement with the conductivity in Fig.~\ref{fig:Conductivity-hub-sigma-xx-yy}(a).

In the case of a partially filled lower band, $\mu<g$, the onset behavior of the conductivity shown in Fig.~\ref{fig:Conductivity-hub-sigma-xx-yy}(b) is also explained by the transitions between $\epsilon_{n=2}$ and $\epsilon_{n=1}$ branches. The corresponding onset frequency $\Omega_{\rm on}\approx 2|g-\mu|$ is determined by the minimal distance between empty states at the $\epsilon_{n=1}$ branch and filled states at the $\epsilon_{n=2}$ branch [see vertical dashed line in Fig.~\ref{fig:Conductivity-hub-sigma-xx-yy}(b)]; in the model at hand, the minimal distance occurs at $k_y=0$. Due to the anisotropic energy spectrum, see Eqs.~(\ref{Model-effective-hub-eps-0})--(\ref{Model-effective-hub-eps-2}), the conductivity does not saturate with $\Omega$. The absence of saturation behavior can be explained by the fact that the whole $\epsilon_{n=2}$ branch cannot contribute to the transitions at any $\hbar \Omega/g\lesssim 1$.

\subsection{Mixed stacking and tilted Dirac model}
\label{sec:App-Conductivity-TDM}

As with the hub-aligned stacking, the interband conductivity for the effective model of the mixed $AA-BC-CB$ stacking depends only on the transitions between the dispersive $\epsilon_{1,2}$ bands. Therefore, we can use the following abbreviated effective Hamiltonian:
\begin{equation}
\label{App-Conductivity-TDM-Heff-2x2}
H_{\rm eff}^{\rm (m)} = \left(g +\frac{\hbar v_F}{2\sqrt{6}}\tilde{k}_x\right) \mathds{1}_2 + \frac{\hbar v_F}{2\sqrt{2}} \left(\tilde{\mathbf{k}}\cdot \bm{\sigma}\right),
\end{equation}
where $\tilde{k}_x = \sqrt{3} k_x$ and $\tilde{k}_y = \sqrt{2} k_y$. The energy spectrum of the above Hamiltonian reads
\begin{equation}
\label{App-Conductivity-TDM-eps-2x2-0}
\epsilon_{\pm} = g + \frac{\hbar v_F}{2\sqrt{6}}\tilde{k}_x \pm \frac{\hbar v_F}{2\sqrt{2}}\tilde{k};
\end{equation}
see also $\epsilon_{n=1,2}$ in Eqs.~(\ref{Model-effective-mixed-eps-0})--(\ref{Model-effective-mixed-eps-2}). As is evident from Eq.~(\ref{App-Conductivity-TDM-eps-2x2-0}), the effective Hamiltonian (\ref{App-Conductivity-TDM-Heff-2x2}) describes a tilted 2D Dirac spectrum~\cite{Xu-Zhang:typeII-2015,Soluyanov-Bernevig:2015,Carbotte:2016-tilt}.

The Green function for the abbreviated Hamiltonian (\ref{App-Conductivity-TDM-Heff-2x2}) reads as
\begin{eqnarray}
\label{App-Conductivity-TDM-G-def}
G(\omega;\mathbf{k}) &=& \frac{1}{\mathcal{D}} \Bigg[\left(\hbar \omega-g-\frac{\hbar v_F}{2\sqrt{6}} \tilde{k}_x\right)\mathds{1}_2 \nonumber\\
&+&\frac{\hbar v_F}{2\sqrt{2}}\left(\tilde{k}_x \sigma_x + \tilde{k}_y\sigma_y\right)\Bigg],
\end{eqnarray}
where the denominator is
\begin{eqnarray}
\label{App-Conductivity-TDM-D-def}
\mathcal{D} &=& \prod_{\eta=\pm}\left(\hbar\omega - \epsilon_{\eta}\right) \nonumber\\
&=& \left[\left(\hbar\omega -g -\frac{\hbar v_F}{2\sqrt{6}}\tilde{k}_x\right)^2 - \frac{(\hbar v_F)^2}{8}\tilde{k}^2 \right].
\end{eqnarray}

The spectral function (\ref{Conductivity-A-def}) is
\begin{eqnarray}
\label{App-Conductivity-TDM-A-def}
A(\omega;\mathbf{k}) &=& \Bigg[\left(\hbar \omega-g-\frac{\hbar v_F}{2\sqrt{6}}\tilde{k}_x\right)\mathds{1}_2  \nonumber\\
&+& \frac{\hbar v_F}{2\sqrt{2}}\left(\tilde{k}_x \sigma_x +\tilde{k}_y\sigma_y\right) \Bigg] \sum_{\eta=\pm} \frac{\delta(\hbar \omega - \epsilon_{\eta})}{\epsilon_{\eta}-\epsilon_{-\eta}}.\nonumber\\
\end{eqnarray}

To calculate the conductivity, we use Eqs.~(\ref{Conductivity-Kubo-sigma-nn}) and (\ref{App-Conductivity-TDM-A-def}). The corresponding traces are
\begin{eqnarray}
\label{App-Conductivity-TDM-vxGvxG}
&&\mbox{tr} \left[ v_x A(\omega; \mathbf{k}) v_x A(\omega-\Omega; \mathbf{k})\right] \nonumber\\
&&= \frac{v_F^2}{96} \Bigg\{96 \left(\hbar \omega -g\right)\left(\hbar \omega -g -\hbar \Omega\right) -(\hbar v_F \tilde{k})^2 +\hbar v_F \tilde{k} \nonumber\\
&&\times \left[4\sqrt{6} \left(2\hbar \omega -2g -\hbar \Omega\right) \cos{(\varphi)}  +5\hbar v_F k \cos{(2\varphi)}\right]
\Bigg\} \nonumber\\
&&\times \sum_{\eta_1,\eta_2=\pm} \frac{\delta(\hbar \omega - \epsilon_{\eta_1})}{\epsilon_{\eta_1}-\epsilon_{-\eta_1}} \frac{\delta(\hbar \omega - \hbar \Omega +\epsilon_{\eta_2})}{\epsilon_{\eta_2}-\epsilon_{-\eta_2}},
\end{eqnarray}
\begin{eqnarray}
\label{App-Conductivity-TDM-vyGvyG}
&&\mbox{tr} \left[ v_y A(\omega; \mathbf{k}) v_y A(\omega-\Omega; \mathbf{k})\right] \nonumber\\
&&= \frac{v_F^2}{96} \Bigg\{48 \left(\hbar \omega -g\right)\left(\hbar \omega -g -\hbar \Omega\right) +(\hbar v_F \tilde{k})^2 -\hbar v_F \tilde{k}\nonumber\\
&& \times \left[4\sqrt{6} \left(2\hbar \omega -2g -\hbar \Omega\right) \cos{(\varphi)}  +5\hbar v_F k \cos{(2\varphi)}\right] \Bigg\} \nonumber\\
&&\times \sum_{\eta_1,\eta_2=\pm} \frac{\delta(\hbar \omega - \epsilon_{\eta_1})}{\epsilon_{\eta_1}-\epsilon_{-\eta_1}} \frac{\delta(\hbar \omega - \hbar \Omega +\epsilon_{\eta_2})}{\epsilon_{\eta_2}-\epsilon_{-\eta_2}},
\end{eqnarray}
where we used the following velocity matrices:
\begin{equation}
\label{App-Conductivity-TDM-vx-vy}
v_x = \frac{v_F}{2\sqrt{6}}\mathds{1}_2 + \frac{v_F}{2\sqrt{2}} \sigma_x, \quad \quad  v_y = \frac{v_F}{2\sqrt{2}} \sigma_y.
\end{equation}

Then, the nontrivial components of the conductivity tensor (\ref{Conductivity-Kubo-sigma-nn}) read
\begin{widetext}
\begin{eqnarray}
\label{App-Conductivity-TDM-sigma-xx}
&&\mbox{Re}{\left\{\sigma_{xx}(\Omega)\right\}} = -\frac{2\pi \hbar v_F^2\sigma_0}{24\sqrt{6}\Omega} \sum_{\eta_1,\eta_2=\pm} \int d^2 \tilde{k} \left[f^{\rm eq}(\epsilon_{\eta_{1}}) -f^{\rm eq}(\epsilon_{\eta_{1}}-\hbar \Omega)\right] \Bigg\{96 \left(\epsilon_{\eta_1} -g\right)\left(\epsilon_{\eta_1} -g -\hbar \Omega\right) -(\hbar v_F \tilde{k})^2 \nonumber\\
&&+\hbar v_F \tilde{k} \left[4\sqrt{6} \left(2\epsilon_{\eta_1} -2g -\hbar \Omega\right) \cos{(\varphi)}  +5\hbar v_F k \cos{(2\varphi)}\right]
\Bigg\} \frac{\delta(\epsilon_{\eta_1} - \hbar \Omega +\epsilon_{\eta_2})}{\left(\epsilon_{\eta_1}-\epsilon_{-\eta_1}\right) \left(\epsilon_{\eta_2}-\epsilon_{-\eta_2}\right)},
\end{eqnarray}
\end{widetext}
\begin{widetext}
\begin{eqnarray}
\label{App-Conductivity-TDM-sigma-yy}
&&\mbox{Re}{\left\{\sigma_{yy}(\Omega)\right\}} = -\frac{2\pi \hbar v_F^2\sigma_0}{24\sqrt{6}\Omega} \sum_{\eta_1,\eta_2=\pm} \int d^2 \tilde{k} \left[f^{\rm eq}(\epsilon_{\eta_{1}}) -f^{\rm eq}(\epsilon_{\eta_{1}}-\hbar \Omega)\right] \Bigg\{48 \left(\epsilon_{\eta_1} -g\right)\left(\epsilon_{\eta_1} -g -\hbar \Omega\right) +(\hbar v_F \tilde{k})^2 \nonumber\\
&&-\hbar v_F \tilde{k} \left[4\sqrt{6} \left(2\epsilon_{\eta_1} -2g -\hbar \Omega\right) \cos{(\varphi)}  +5\hbar v_F k \cos{(2\varphi)}\right] \Bigg\}
\frac{\delta(\epsilon_{\eta_1} - \hbar \Omega +\epsilon_{\eta_2})}{\left(\epsilon_{\eta_1}-\epsilon_{-\eta_1}\right) \left(\epsilon_{\eta_2}-\epsilon_{-\eta_2}\right)}.
\end{eqnarray}
\end{widetext}

Since we are interested in the interband transitions $\eta_1 = -\eta_2$, we rewrite the $\delta$ functions in the above equation as
\begin{eqnarray}
\label{App-Conductivity-TDM-delta}
&&\delta{\left(\epsilon_{\eta_1} - \hbar \Omega +\epsilon_{\eta_2}\right)} = \delta{\left((\eta_1-\eta_2) \frac{\hbar v_F}{2\sqrt{2}}\tilde{k} -\hbar \Omega\right)} \nonumber\\
&&\stackrel{\eta_1 = -\eta_2}{=}
\frac{\sqrt{2}}{\hbar v_F} \delta{\left(\tilde{k} - \frac{\sqrt{2} \Omega}{v_F}\right)}.
\end{eqnarray}
The $\delta$ function given in Eq.~(\ref{App-Conductivity-TDM-delta}) allows us to integrate over $\tilde{k}$ in Eqs.~(\ref{App-Conductivity-TDM-sigma-xx}) and (\ref{App-Conductivity-TDM-sigma-yy}):
\begin{widetext}
\begin{eqnarray}
\label{App-Conductivity-TDM-sigma-xx-1}
&&\mbox{Re}{\left\{\sigma_{xx}(\Omega)\right\}} = -\frac{20 \sigma_0}{19\sqrt{6}} \int_{0}^{2\pi} \frac{d\varphi}{2\pi} \left[f^{\rm eq}\left(\frac{\hbar \Omega}{2} + \frac{\cos{(\varphi)}}{\sqrt{3}}\right) -f^{\rm eq}\left(-\frac{\hbar \Omega}{2} + \frac{\cos{(\varphi)}}{\sqrt{3}}\right)\right] \left[\frac{26}{10} -\cos{(2\varphi)}\right],\\
\label{App-Conductivity-TDM-sigma-yy-1}
&&\mbox{Re}{\left\{\sigma_{yy}(\Omega)\right\}} = -\frac{20 \sigma_0}{19\sqrt{6}} \int_{0}^{2\pi} \frac{d\varphi}{2\pi} \left[f^{\rm eq}\left(\frac{\hbar \Omega}{2} + \frac{\cos{(\varphi)}}{\sqrt{3}}\right) -f^{\rm eq}\left(-\frac{\hbar \Omega}{2} + \frac{\cos{(\varphi)}}{\sqrt{3}}\right)\right] \left[1 +\cos{(2\varphi)}\right].
\end{eqnarray}
\end{widetext}

Similar to the hub-aligned $AB-BA-CC$ stacking, the isotropy of the conductivity is restored after we average over all equivalent pairs of the crossing points; the resulting conductivity is then given by $(\sigma_{xx}+\sigma_{yy})/2$.

\begin{figure*}[!ht]
\centering
\subfigure[]{\includegraphics[width=0.45\textwidth]{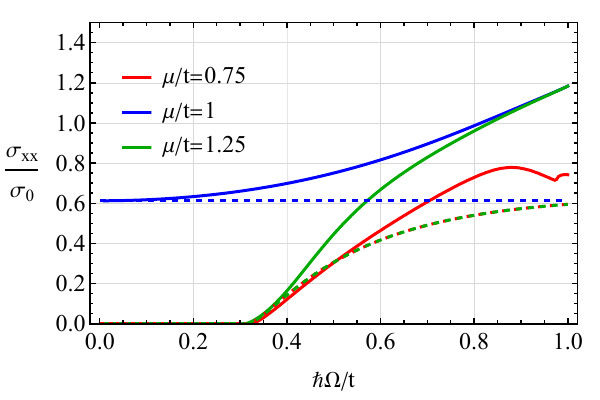}}
\hspace{0.01\textwidth}
\subfigure[]{\includegraphics[width=0.45\textwidth]{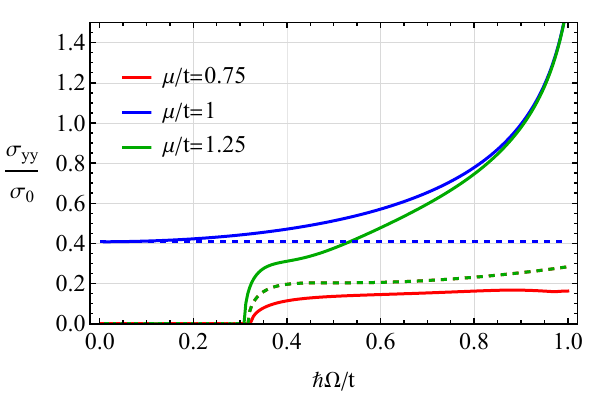}}
\caption{
The dependence of the interband conductivity tensor components $\RE{\sigma_{xx}}$ (panel (a)) and $\RE{\sigma_{yy}}$ (panel (b)) normalized to $\sigma_0 = e^2/(4\hbar)$ on $\hbar\Omega/t$ for the mixed $AA-BC-CB$ stacking at a few values of $\mu/t$. Solid and dashed lines correspond to the non-abbreviated and abbreviated effective models, respectively; see Eqs.~(\ref{App-Model-effective-hub-Heff}) and (\ref{Conductivity-hub-SD-H-def}), respectively. In all panels, we set $T\to0$ and $g/t=1$.
}
\label{fig:Conductivity-mixed-sigma-xx-yy}
\end{figure*}

We present the $xx$ and $yy$ components of the optical conductivity tensor in Fig.~\ref{fig:Conductivity-mixed-sigma-xx-yy} for the nonabbreviated and abbreviated effective models; see Eqs.~(\ref{App-Model-effective-mixed-Heff}) and (\ref{App-Conductivity-TDM-Heff-2x2}), respectively. As one can see, while the non-abbreviated effective model has a quantitatively different profile of optical conductivity and is particle-hole asymmetric, the key features, e.g., the onset frequencies, agree well with those in the abbreviated model.

\subsection{Cyclic \texorpdfstring{$AB-BC-CA$}{AB-BC-CA} stacking}
\label{sec:App-Conductivity-cyclic}

Finally, we discuss the optical conductivity for the effective model of the cyclic $AB-BC-CA$ stacking; see Eq.~(\ref{Model-effective-cyclic-Heff}) for the corresponding effective Hamiltonian. We have the following velocity matrices:
\begin{eqnarray}
\label{App-Conductivity-cyclic-vx}
v_x &=& \frac{v_F}{\sqrt{2}} \left(
                                 \begin{array}{ccc}
                                   0 & 1/2 & 1/2 \\
                                   1/2 & 0 & 1 \\
                                   1/2 & 1 & 0 \\
                                 \end{array}
                               \right), \nonumber\\
v_y &=& \frac{v_F}{\sqrt{2}} \left(
                                 \begin{array}{ccc}
                                   0 & -i/2 & i/2 \\
                                   i/2 & 0 & -i \\
                                   -i/2 & i & 0 \\
                                 \end{array}
                               \right).
\end{eqnarray}

The expressions for the Green function and the spectral function can be straightforwardly obtained but are bulky. Therefore, we do not present them here. The traces in the conductivity defined in Eq.~(\ref{Conductivity-Kubo-sigma-nn}) read
\begin{widetext}
\begin{eqnarray}
\label{App-Conductivity-cyclic-vxAvxA}
&&\mbox{Tr}{\left(v_x A(\omega; \mathbf{k}) v_x A(\omega-\Omega; \mathbf{k}) \right)} =
F(\omega) F(\omega -\Omega) v_F^2 \frac{3}{32}
\Bigg\{16\left(g-\hbar \omega \right)^2 \left(g-\hbar \omega +\hbar \Omega\right)^2 \nonumber\\
&&-4\sqrt{2} \hbar v_F k \left(g-\hbar \omega \right) \left(2g-2\hbar \omega +\hbar \Omega\right) \left(g-\hbar \omega +\hbar \Omega\right) \cos{(\varphi)}
-3(\hbar v_F k)^2 \left[(\hbar \Omega)^2 -\left(2g-2\hbar \omega +\hbar \Omega\right)^2 \cos{(2\varphi)}\right] \nonumber\\
&&-2\sqrt{2} (\hbar v_F k)^3 \left(2g-2\hbar \omega +\hbar \Omega\right) \cos{(3\varphi)}
+(\hbar v_F k)^4 \left[\cos^4{(\varphi)} + 3\sin^4{(\varphi)}\right]
\Bigg\},
\end{eqnarray}
\end{widetext}
\begin{widetext}
\begin{eqnarray}
\label{App-Conductivity-cyclic-vyAvyA}
&&\mbox{Tr}{\left(v_y A(\omega; \mathbf{k}) v_y A(\omega-\Omega; \mathbf{k}) \right)} =
F(\omega) F(\omega -\Omega) v_F^2 \frac{3}{32}
\Bigg\{16\left(g-\hbar \omega \right)^2 \left(g-\hbar \omega +\hbar \Omega\right)^2 \nonumber\\
&&+4\sqrt{2} \hbar v_F k \left(g-\hbar \omega \right) \left(2g-2\hbar \omega +\hbar \Omega\right) \left(g-\hbar \omega +\hbar \Omega\right) \cos{(\varphi)}
-3(\hbar v_F k)^2 \left[(\hbar \Omega)^2 +\left(2g-2\hbar \omega +\hbar \Omega\right)^2 \cos{(2\varphi)}\right] \nonumber\\
&&-2\sqrt{2} (\hbar v_F k)^3 \left(2g-2\hbar \omega +\hbar \Omega\right) \cos{(3\varphi)}
+2(\hbar v_F k)^4 \cos^2{(\varphi)}\left[2-\cos{(2\varphi)}\right]
\Bigg\}.
\end{eqnarray}
\end{widetext}
Here, $F(\omega)$ is defined in Eq.~(\ref{App-Conductivity-aligned-F-def}) with the energy spectrum given in Eq.~(\ref{Model-effective-cyclic-eps-0-1-2}). In order to calculate the conductivity, we rewrite
\begin{equation}
\label{App-Conductivity-cyclic-delta}
\delta(\epsilon_{n_1} -\epsilon_{n_2} -\hbar \Omega) = \frac{1}{\hbar \left|\Delta_{n_1,n_2}\right|} \delta\left(v_F k -\frac{\Omega}{\Delta_{n_1,n_2}}\right),
\end{equation}
where
\begin{equation}
\label{Conductivity-cyclic-omega-nn}
\Delta_{n_1,n_2} = \frac{1}{\hbar v_F}\partial_k\left(\epsilon_{n_1}-\epsilon_{n_2}\right)
\end{equation}
and the energy dispersion $\epsilon_{n}$ is given in Eq.~(\ref{Model-effective-cyclic-eps-0-1-2}).
This expression enters Eq.~(\ref{App-Conductivity-aligned-F-def}) and allows us to straightforwardly integrate over $k$ in the interband terms of the conductivity. By using Eqs.~(\ref{App-Conductivity-aligned-F-def}), (\ref{App-Conductivity-cyclic-vxAvxA}), and (\ref{App-Conductivity-cyclic-delta}) in Eq.~(\ref{Conductivity-Kubo-sigma-nn}), we obtain the diagonal components of the real part of the interband conductivity:
\begin{widetext}
\begin{eqnarray}
\label{Conductivity-cyclic-sigma-xx}
\RE{\sigma_{xx}(\Omega)} &=& \frac{3 \sigma_0}{8}
\sum_{n_1, n_2=0}^2\int_{0}^{2\pi} \frac{d\varphi}{2\pi} \frac{\theta(\Delta_{n_1,n_2})}{|\Delta_{n_1,n_2}|^2} \frac{f^{\rm eq}(\epsilon_{n_1}-\hbar\Omega) -f^{\rm eq}(\epsilon_{n_1})}{\Pi_{m_1=0}^{2, \prime} (\epsilon_{n_1}-\epsilon_{m_1}) \Pi_{m_2=0}^{2, \prime} (\epsilon_{n_2}-\epsilon_{m_2})}
\Bigg\{16\left(g -\epsilon_{n_1} \right)^2 \left(g -\epsilon_{n_1} +\hbar \Omega\right)^2 \nonumber\\
&-& 4\sqrt{2} \frac{\hbar \Omega}{\Delta_{n_1,n_2}} \left(g-\epsilon_{n_1} \right) \left(2g-2\epsilon_{n_1} +\hbar \Omega\right) \left(g-\epsilon_{n_1} +\hbar \Omega\right) \cos{(\varphi)} \nonumber\\
&-& 3\left(\frac{\hbar \Omega}{\Delta_{n_1,n_2}}\right)^2 \left[(\hbar \Omega)^2 -\left(2g-2\epsilon_{n_1} +\hbar \Omega\right)^2 \cos{(2\varphi)}\right] \nonumber\\
&-& 2\sqrt{2} \left(\frac{\hbar \Omega}{\Delta_{n_1,n_2}}\right)^3 \left(2g-2\epsilon_{n_1} +\hbar \Omega\right) \cos{(3\varphi)} +\left(\frac{\hbar \Omega}{\Delta_{n_1,n_2}}\right)^4 \left[\cos^4{(\varphi)} + 3\sin^4{(\varphi)}\right]
\Bigg\}.
\end{eqnarray}
\end{widetext}
It can be shown that $\RE{\sigma_{xx}} =\RE{\sigma_{yy}}$ and, as expected, $\RE{\sigma_{xy}}=0$.

\begin{figure*}[!ht]
\centering
\subfigure[]{\includegraphics[width=0.45\textwidth]{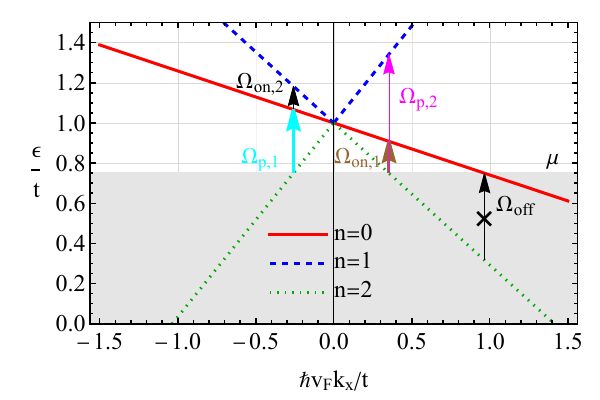}}
\hspace{0.01\textwidth}
\subfigure[]{\includegraphics[width=0.45\textwidth]{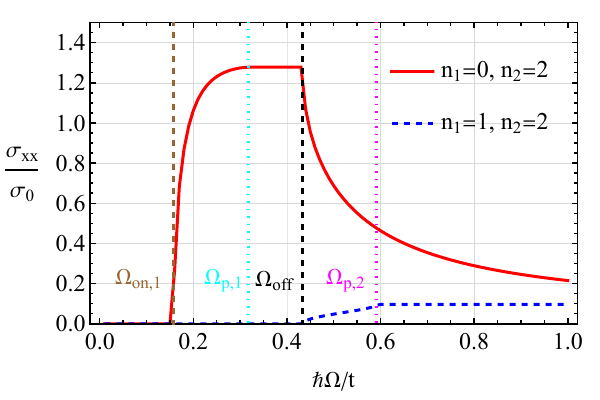}}
\caption{
The cross-section of the energy spectrum at $k_y=0$ for the cyclic $AB-BC-CA$ stacking and a few possible transitions at $\mu=0.75\,t$ are shown in panel (a). The frequencies of these transitions discussed in the main text are shown by arrows. The contributions to the interband conductivity in the effective model from the allowed transitions at $\mu=0.75\,t$ are shown in panel (b) at $T\to0$. In both panels, we fixed $g/t=1$.
}
\label{fig:Conductivity-cyclic-sigma-effective-1}
\end{figure*}

To explain the dependence of the interband part of the conductivity on frequency shown in Fig.~\ref{fig:Conductivity-cyclic-sigma-effective-num}, we investigate the activation behavior of each of the transitions between three different branches of the effective model; unlike the hub-aligned and mixed stackings, all bands should be taken into account for the cyclic stacking. We present the dispersion relation of the effective Hamiltonian, see Eq.~(\ref{Model-effective-cyclic-eps-0-1-2}), at $k_y=0$ and $g/t=1$ in Fig.~\ref{fig:Conductivity-cyclic-sigma-effective-1}(a); for definiteness, we fix $\mu=0.75\,t$. The contributions to the conductivity from different transitions are shown in Fig.~\ref{fig:Conductivity-cyclic-sigma-effective-1}(b). As one can see from Fig.~\ref{fig:Conductivity-cyclic-sigma-effective-1}(a), the $\epsilon_{n=0}$ branch is not flat. Therefore, the transitions between $\epsilon_{n=0}$ and $\epsilon_{n=2}$ branches are allowed even for $\Omega > \Omega_{\rm on, 1}$, where $\Omega_{\rm on, 1} < g-\mu$. Here, $\hbar \Omega_{\rm on, 1}/t \approx 0.16$ is determined by the minimal distance between occupied $\epsilon_{n=2}$ and empty $\epsilon_{n=0}$ branches; see Fig.~\ref{fig:Conductivity-cyclic-sigma-effective-1}(a) and the onset of the plateau in Fig.~\ref{fig:Conductivity-cyclic-sigma-effective-1}(b) marked by a thick vertical dashed black line. The conductivity in Fig.~\ref{fig:Conductivity-cyclic-sigma-effective-1}(b) saturates at $\hbar \Omega_{\rm p, 1}/t \approx 0.32$ determined by the condition that the whole $\epsilon_{n=2}$ branch can contribute to the optical conductivity. For larger frequencies, $\hbar \Omega/t > \hbar \Omega_{\rm off}/t \approx 0.43$, we observe a decrease of the conductivity explained by the fact that only a part of the $\epsilon_{n=2}$ branch can contribute to the transitions between the $\epsilon_{n=2}$ and $\epsilon_{n=0}$ branches due to the Pauli blocking; see Fig.~\ref{fig:Conductivity-cyclic-sigma-effective-1}(a). The offset frequency $\Omega_{\rm off}$ corresponds to the minimal distance between the filled parts of $\epsilon_{n=2}$ and $\epsilon_{n=0}$ branches. At the same frequency $\Omega_{\rm on,2}\approx \Omega_{\rm off}$, the transitions between $\epsilon_{n=2}$ and $\epsilon_{n=1}$ branches become possible, which is manifested as a relatively small contribution to the conductivity; see the blue dashed line in Fig.~\ref{fig:Conductivity-cyclic-sigma-effective-1}(b). This contribution saturates at $\hbar \Omega_{\rm p,2}/t \approx 0.59$ for which the whole $n=2$ branch can contribute.

\bibliography{library-short}

\end{document}